\documentclass[useAMS,usenatbib]{mn2e}

\usepackage{graphicx}

\title[Integral Field Spectroscopy of Local LCBGs: NGC 7673, a case
study.]{Integral Field Spectroscopy of Local LCBGs: NGC 7673, a case study. Physical properties of star-forming regions}
\author[A. Castillo-Morales, et al.] {A. Castillo-Morales$^{1}$\thanks{E-mail:acm@astrax.fis.ucm.es}, J. Gallego$^{1}$, J. P\'erez-Gallego$^{2}$, R. Guzm\'an$^{2}$, J.C. Mu\~{n}oz-Mateos$^{1,4}$,
\newauthor
 J. Zamorano$^{1}$ and S.F. S\'anchez$^{3}$\\
$^{1}$Dpto. de Astrof\'{\i}sica y CC. de la Atm\'osfera, Universidad Complutense de Madrid, Spain.\\
$^{2}$University of Florida, 211 Bryant Space Science Center,FL 32611-2055, Gainesville, USA. \\
$^{3}$Centro Astron\'omico Hispano Alem\'an, AIE (CSIC-MPG, E-04004 Almer\'ia, Spain\\
$^{4}$National Radio Astronomy Observatory, 520 Edgemont Road, Charlottesville, VA 22903-2475, USA}
\begin{document}

\maketitle

\label{firstpage}

\begin{abstract}

Physical properties of the star-forming regions in the local Luminous Compact Blue Galaxy (LCBG) NGC 7673 are studied in detail using 3D spectroscopic data taken with the PPAK integral field unit at the 3.5-m telescope in the Centro Astron\'omico Hispano Alem\'an.
We derive integrated and spatially resolved properties such as extinction, star formation rate (SFR) and metallicity for this galaxy. 
Our data show an extinction map with maximum values located at the position of the main clumps of star formation showing small spatial variations ($\rm E(B-V)_{t}=0.12-0.21 mag$).
We derive a $\rm H\alpha$-based SFR for this galaxy of $6.2 \pm 0.8 \rm M_{\odot}/yr$ in agreement with the SFR derived from infrared and radio continuum fluxes.
The star formation is located mainly in clumps A, B, C and F. 
Different properties measured in clump B makes this region peculiar. We find the highest $\rm H\alpha$ luminosity with a SFR surface density of 0.5 $\rm M_{\odot}yr^{-1}kpc^{-2}$ in this clump.
In our previous work \citep{perez-gallego10}, the kinematic analysis for this galaxy shows an asymmetrical ionized gas velocity field with a kinematic decoupled component located at the position of clump B. This region shows the absence of strong absorption features and the presence of a Wolf-Rayet stellar population indicating this is a young burst of massive stars. 
Furthermore, we estimate a gas metallicity of $\rm 12+log(O/H)=8.20\pm0.15$ (0.32 solar) for the integrated galaxy using the R23 index. The values derived for the different clumps with this method show small metallicity variations in this galaxy, with values in the range 8.12 (for clump A) - 8.23 (for clump B) for $\rm 12+log(O/H)$.
The analysis of the emission line ratios discards the presence of any AGN activity or shocks as the ionization source in this galaxy.
Between the possible mechanisms to explain the starburst activity in this galaxy, our 3D spectroscopic data support the scenario of an on-going interaction with the possibility for clump B to be the dwarf satellite galaxy.

\end{abstract}


%% subject header.  Objects should be in the appropriate "individual"
%% headers (e.g. quasars: individual, stars: individual, etc.) with the
%% additional provision that the total number of headers, including each
%% individual object, not exceed six.  The \objectname{} macro, and its
%% alias \object{}, is used to mark each object.  The macro takes the object
%% name as its primary argument.  This name will appear in the paper
%% and serve as the link's anchor in the electronic edition if the name
%% is recognized by the data centers.  The macro also takes an optional
%% argument in parentheses in cases where the data center identification
%% differs from what is to be printed in the paper.


\section{Introduction}

Starburst galaxies are those in which star formation and associated phenomena dominate the total energetics \citep{weedman83}. These galaxies have larger star formation rates (SFRs) per unit area than normal galaxies, and to produce their current stellar masses at their current SFRs they would take much less than the age of the Universe \citep{kennicutt98a}. Starburst galaxies are found at different redshifts, which denotes their cosmological relevance, and turn those found nearby into perfect candidates to study the mysteries of the star formation process throughout time when they can be equally and properly selected at different epochs of the Universe. Both distant Lyman Break Galaxies \citep[LBGs,][]{steidel96,lowenthal97} and closer Luminous Compact Blue Galaxies \citep[LCBGs,][]{werk04} fall into this category.

LCBGs are, as described by \citet{werk04}, galaxies with (i) absolute blue magnitude ($M_B$) brighter than -18.5; (ii) effective surface brightness ($SB_e$) brighter than 21 $B$-mag arcsec$^{-2}$; and (iii)($B-V$) colour bluer than 0.6. In an observational parameter space defined by these observational properties (i.e., $M_B$, $SB_{e}$, and ($B-V$)) LCBGs and distant LBGs share the same optical properties.
Galaxies belonging to this population (i) are morphologically heterogeneous, (ii) form stars at around 10 \--- 20 $M_{\odot}$ yr$^{-1}$, (iii) show velocity widths of 30 \--- 120 km s$^{-1}$, (iv) are as compact as 2 \--- 5 kpc, and (v) have metallicities lower than solar.

LCBGs play an important role in galaxy evolution over cosmological time scales, as has been shown by various observational studies in the last decade \citep{werk04, melbourne07, guzman97,steidel03}. The lack of LCBGs in the local universe when compared to higher redshifts arises the issue of how this population has evolved in the last 9 \--- 10 Gyr. Two are the main suggested scenarios (not mutually exclusive): (i) LCBGs are the progenitors of today's spheroidal galaxies, low-mass ($M<10^{10}$ $M_{\odot}$) spheroidal galaxies or dwarf elliptical galaxies whose properties according to evolutionary models would be matched by a typical LCBG after a 4 \--- 6 Gyr fading process showing low metallicities \citep{koo94,guzman98,noeske06}; and (ii) LCBGs are the progenitors of the spheroidal component of today's disk galaxies, present day small spirals, more massive ($M\sim10^{10}$ $M_{\odot}$) than inferred from virial masses, whose emission is mostly due to a vigorous central burst \citep{phillips97,hammer01,puech06}. This scenario predicts higher metallicities than those likely to be observed in local dwarf galaxies \citep{kobulnicky99}.

Key ingredients on the discussion on whether LCBGs evolve one way or another is their mass and metallicity. 
A reliable determination of their masses and metallicities is necessary to properly place LCBGs within one evolution scenario or the other, or to understand what makes them evolve one way or another.
Masses of LCBGs can be derived from their rotation curves and velocity widths, nevertheless one needs to be careful with those, since most of these objects' kinematics might not be coupled to their masses due to supernova galactic winds, and both minor and major mergers. 
The metallicity can be estimated in LCBGs by measuring the auroral emission line flux $\rm [O III]\lambda4363$, which gives, along with the $\rm [O III]\lambda\lambda4959, 5007$ fluxes, a direct determination of the electron temperature. But beyond certain metallicities $\rm [O III]\lambda4363$ detection is not possible and the oxygen abundance must be empirically determined using strong emission lines through the use, for example, of the R23 indicator. 

The knowledge not only of the integrated properties such as mass, metallicity, extinction or star-formation rate, but their spatial distribution in the galaxy plays an important role to study the nature of LCBGs.  
Integral field spectroscopy technique offers a great advantage to carry out this issue. In this way, we are able to derive full maps of physical properties such as extinction to locate the dusty regions in these galaxies. This is important to properly compute the extinction-corrected emission line maps and derive star formation rate and metallicity spatial distributions. 3D spectroscopy makes it possible to carry out this work in an efficient way allowing us to test model predictions on the origin and evolution of these massive starbursts.

Interest in LCBGs has multiplied following the initial observational results and the properties highlighted above, but their relation to today's galaxy population still remains unknown. In order to understand the nature of LCBGs and its role in galaxy evolution we have selected a representative sample of 22 LCBGs within 200 Mpc from the Sloan Digital Sky Survey~\citep[SDSS,][]{adelman06}, Universidad Complutense de Madrid~\citep[UCM,][]{zamorano94} and Markarian catalogs~\citep{markarian89} that best resemble the properties of distant LCBGs, ensuring that this sample, although small, is representative of the LCBG population as a class by covering the whole range in luminosity, colour, surface brightness, and environment (see~\citet{perez-gallego10} (hereafter PG10) for more details on the sample selection). We are carrying out a multi-wavelength study of this sample including not only the optical, but also the millimeter and centimeter ranges \citep[e.g.,][]{garland07}. Arguably, one of the most important aspects of this study and the one we focus on this paper is the optical three dimensional (3D) spectroscopy. The optical is the best understood spectral range in nearby galaxies, and will be systematically studied at high-$z$ with the new generation of near-infrared multi-object spectrographs and integral field units in 10-m class telescopes \citep{forster06,puech06}. 

In this paper we focus on the physical properties of the star-forming regions of LCBG NGC 7673. We use integral field spectroscopy data which allow us to study both the integrated and the spatially resolved properties, such as extinction, SFR and gas metallicity. The detailed analysis of these properties together with the kinematic analysis of the ionized gas discussed in~PG10, will help to understand the starburst nature of this galaxy.

NGC 7673, also known as UCM2325+2318 and MRK325, is a nearby ($z = 0.011368$) starburst
galaxy widely studied in the past \citep[e.g.,][]{duflot82,homeier02,pasquali08,homeier99}. The small size, high surface brightness, strong emission lines and blue colours make NGC 7673 a prototypical LCBG (see Table~\ref{table:NGC7673properties}).

The starburst activity in this galaxy is located within a circle of 3.7 kpc radius ($\rm \approx 2\times R_{e}$, {where $R_{e}$ is the effective radius). It shows a clumpy structure with star-forming regions visible in optical as bright knots in the
galactic disk \citep[][detected 87 distinct star-forming
knots in a deep $\rm H\alpha$ image of this galaxy]{perez-gonzalez03}. \cite{homeier02} identified 50 star cluster candidates working on HST/WFPC2 images, where the bluer and brighter clusters are strongly concentrated into the main starburst ``clumps'' (see Figure~\ref{figure:hst_not}). 
From previous ground-based images~\citep{duflot82} six main clumps have been identified A,B,C,D,E,F (see their Figure 1) and will be referred to in this paper. 

The cosmology used in this paper is $\rm H_{0}$ = 70 km s$^{-1}$ Mpc$^{-1}$. The recession velocity of 3408~km/s from NASA/IPAC Extragalactic Database, implies a distance of $\rm d = 49~Mpc$, and a projected scale of 250 pc per arcsec for this galaxy. This paper is structured as follows. Observations and data reduction are described in Section 2. Our measurements are shown in Section 3. The results and discussion are carried out in Section 4 and in Section 5. Finally, Section 6 shows the summary of our work.

\begin{table}
\centering
\caption{NGC 7673 Observational Properties.  Redshift z from \protect \cite{huchra99}. Absolute B magnitude M(B),  effective surface brightness $\rm SB_{e}$, colour index (B-V) and effective radius $R_{e}$ from \protect  \cite{pisano01} .}
\begin{tabular}{cc}
\hline
\hline
    Name &  NGC 7673 \\
    z &  0.011368\\  
    M(B) & -20.50 mag \\
    $SB_{e} $& 19.40 mag $\rm arcsec^{2}$\\
    (B-V)  & 0.30 mag \\
    $R_{e}$ & 1.9 kpc  \\
\hline
\end{tabular}
\label{table:NGC7673properties}
\end{table}

\section{Observations and Data Reduction}
\label{section:datareduction}

Objects from our sample were observed using PPAK \citep{kelz06}, a
fibre based system for integral field spectroscopy operating at the 3.5-m
CAHA telescope \footnote{Based on observations collected at the German-Spanish Astro-nomical Center, Calar Alto, jointly operated by the Max-Planck-Institut f\"{u}r Astronomie Heidelberg and the Instituto de Astrof\'{i}sica de Andaluc\'{i}a (IAA/CSIC).}. The field of view of PPAK, as well as its good
spatial and spectral resolution is optimum for the observation of our
sample galaxies with typical half-light diameters of around $10^{''}$
to $30^{''}$, and velocity widths ranging from 60 to 120 km s$^{-1}$. PPAK
consists of 331 scientific fibres each of $2.7^{''}$ in diameter,
covering an hexagonal area of $74^{''}$ x $65^{''}$ on the sky. In
addition, there are 15 calibration fibres and 36 fibres grouped in 6 bundles located at $\approx 90^{''}$ of the centre of the science bundle that are used to measure the sky simultaneously.

PPAK observations of NGC 7673 were made over the nights of 2005 August
10 and 11 using two different setups. Firstly, a 300 lines mm$^{-1}$
grating (V300) centred at 5316 {\AA} was used. This low resolution
configuration provided a spectral resolution of 10.7 {\AA} FWHM
covering from 3600 to 7000 {\AA} and allowed us to measure all the
emission line ratios in one single spectrum, avoiding uncertainties
associated to flux calibration and spectral response between different
spectral ranges obtained with different configurations. 
Secondly, a 1200 lines mm$^{-1}$ grating (V1200) centred at 5040
{\AA} was used. This intermediate resolution configuration provided a
nominal spectral resolution of 2.78 {\AA} FWHM (i.e. $\sigma\sim75$ km
s$^{-1}$ at H$\beta$), covering from 4900 to 5400 {\AA}. 
Three different dithering positions were observed in both spectral configurations (see Table 2 in~PG10 for more observational details.) 

The data reduction of two-dimensional fibre spectra was done using the IRAF\footnote{IRAF is distributed by the National Optical Astronomy Observatories} environment, R3D and Euro3D software~\citep{sanchez04} following the techniques described in \cite{sanchez06}. The standard procedure consists on the following steps: bias subtraction, spectra extraction, wavelength calibration (see PG10 for more data reduction details in this step) , fibre to fibre correction and sky emission subtraction.
                                
 The aforementioned data reduction steps are done for each individual
dithering exposure. The following steps instead are done using the information of the 3 dithering exposures to compute a final data cube absolute flux calibrated.

{\bf{ (i) Flux calibration}}. PPAK is based on pure fibre-bundles and does not
cover the entire field-of-view, which imposes flux losses. Nevertheless our observational technique of dithering allows us to cover the complete field of view in three exposures. Therefore,
our first approach is to determine a relative spectrophotometric calibration for each dithering exposure and
recalibrate the spectra later, using additional information coming from broad-band photometry. The relative flux calibration requires the observation of a standard spectrophotometric calibration star. We apply all the previous reduction steps to the calibration star frame. We compare
the measured spectrum of the star (flux coming from several fibres)
with the absolute values and determine the ratio between counts per second and flux. In this way we get the instrumental
response which is applied to science frames to flux-calibrate them.
To perform an absolute flux calibration we compute broad-band PPAK maps in B, V and R filters. For B and R broad-band maps, fluxes obtained in different circular apertures over the PPAK field-of-view are compared with fluxes computed over JKT (B-band) and NOT (R-band) Telescopes at La Palma, from~\cite{perez-gonzalez00, perez-gonzalez03} with the same apertures. JKT and NOT images were registered taking as a reference the PPAK map and then convolved with a Gaussian filter to match the spatial resolution of PPAK. For V-band filter the \cite{huchra77} tabulated value, $\rm m_{V}$=13.11, is compared with the apparent magnitude computed from the PPAK V-band map. 
The comparison of relative PPAK fluxes with absolute fluxes in different photometric bands and in different apertures gives a mean factor of 0.44 with a standard deviation of 0.04, i.e PPAK fluxes need to be multiplied by this factor to match absolute fluxes.
We compared the absolute PPAK H$\alpha$ emission fluxes computed in different apertures with the fluxes from the H$\alpha$ narrow-band image from~\cite{perez-gonzalez03}. In this case PPAK fluxes are $\approx 15\% $ smaller compared with those in the H$\alpha$ narrow-band image.

{\bf{(ii) Locating the spectra in the sky}}. The location of the spectra in
the sky is given by a position table that relates each spectrum with a
certain fibre. 
Since we use the observational technique of dithering
exposures in order to cover the complete field-of-view, the next step
in the data reduction is to create a data cube with the information for
the three ditherings. 
This is done through the $regularization$ of the
data: a grid of 1 arcsec/pixel is constructed over the field of view of the dithered exposures. Then, for each pixel of this grid, the spectra that covers the pixel are averaged (weighted by the
fraction of pixel area in each dithering), obtaining a final
data cube. This $regularization$ does not imply any interpolation. Finally, we obtain a data cube with a spectrum in each pixel where the exposure time for each pixel is different, depending on whether it has been covered by one, two or three fibres.

{\bf{(iii) Differential Atmospheric Refraction (DAR) correction}}. This
correction is done by tracing the location of the intensity peak of a
reference object in the field-of-view along the spectral range. These
locations are estimated by determining the centroid of the object in
the image slice extracted at each wavelength from the data cube. Then,
the full data cube is shifted to a common reference by re-sampling and
shifting each image slice at each wavelength using an interpolation
scheme in the spatial direction. In the case of V300 configuration we apply maximum shifts of 0.2 and 0.4 arcsec in the x and y axis, respectively. For V1200 configuration we checked DAR correction is not required with shifts smaller than 0.1 arcsec.

%---------------------------------------------
\section{Data Measurements}
%---------------------------------------------

\begin{figure}
\center
\includegraphics[height=6.cm]{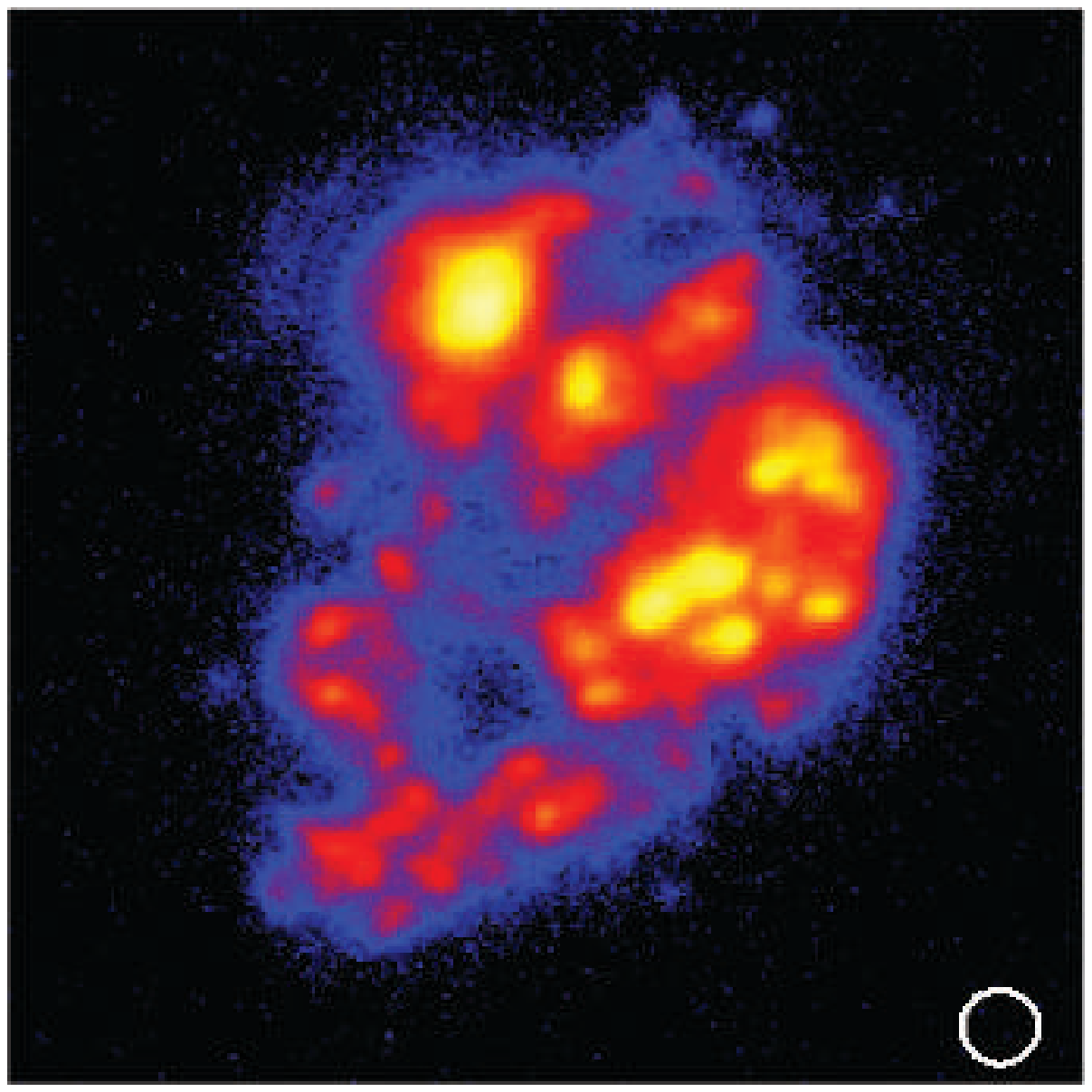}
\includegraphics[height=6.cm]{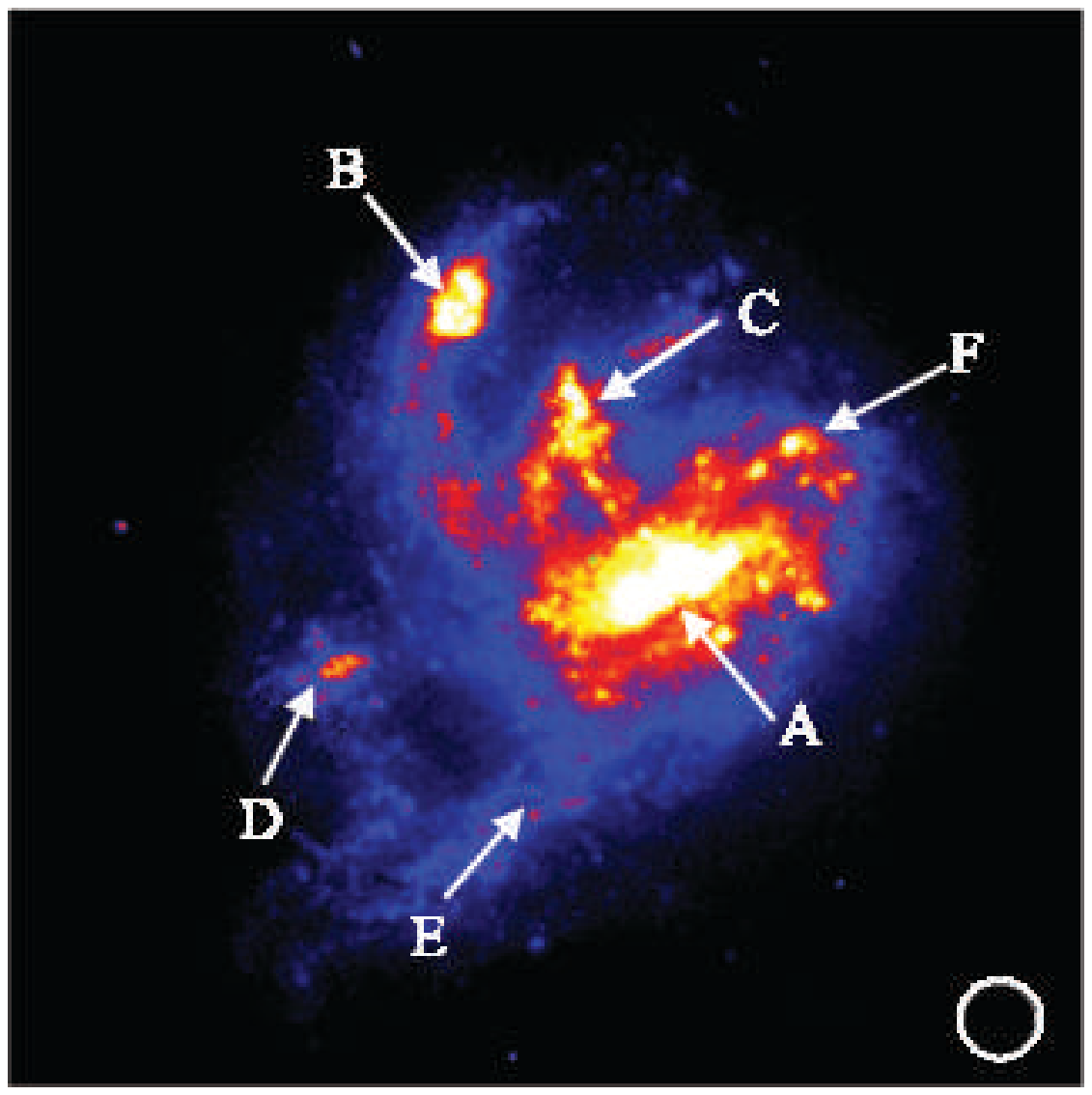}
\caption{Top: $\rm H\alpha$ image from the NOT Telescope (\protect \cite{perez-gonzalez03}). Bottom: Optical image (WFPC2, F555W filter) from HST. The positions of the main clumps A to F are indicated. Both images are 40''x40'' and the PPAK spaxel (2.7'' diameter) is plotted at the right bottom corner.}
\label{figure:hst_not}     
\end{figure}

\begin{figure}
\center
\includegraphics[height=6.cm]{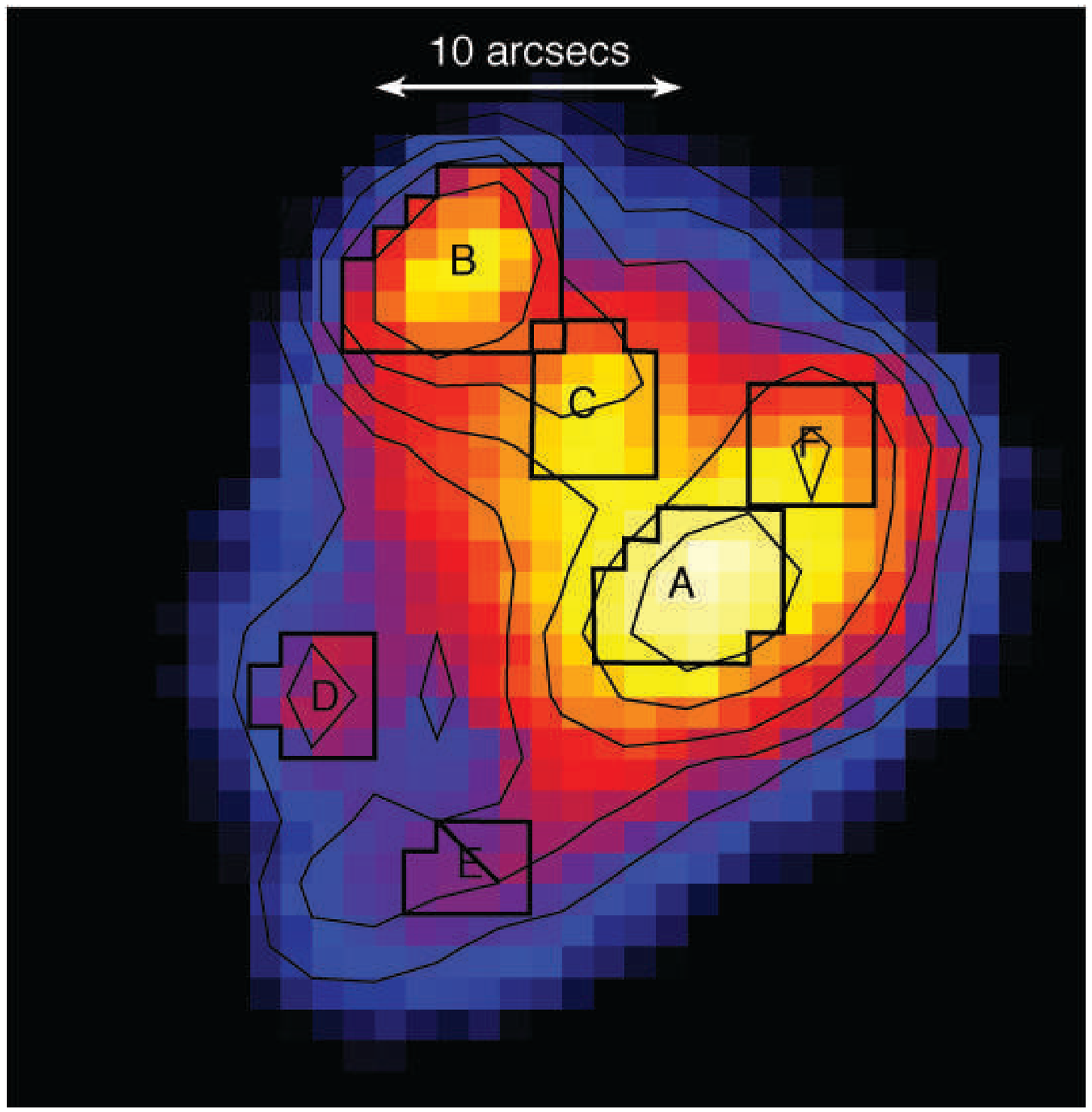}
\includegraphics[height=6.cm]{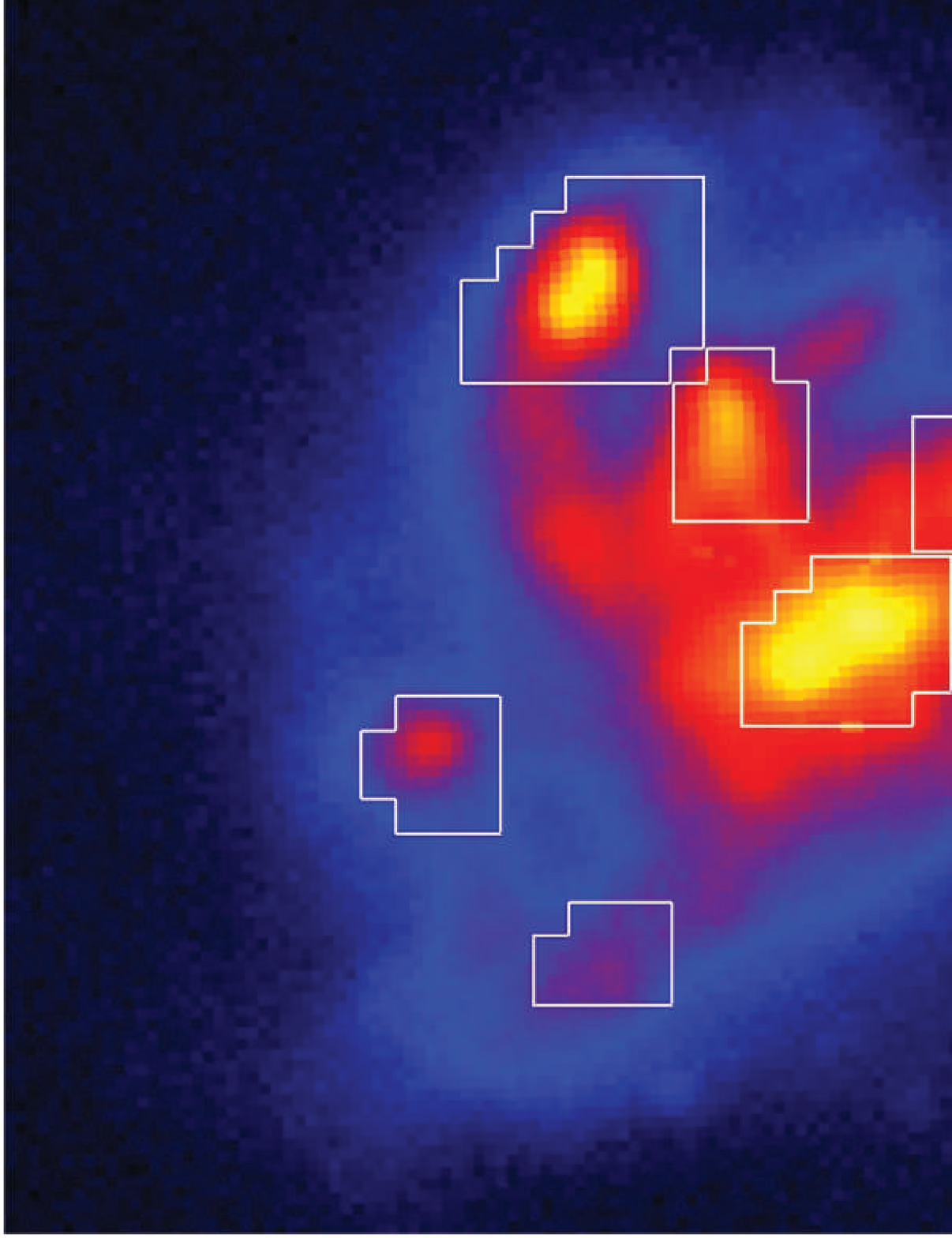}
\caption{Top: PPAK B-band continuum map and $\rm H\alpha$ emission overlaid in contours. The positions of the main clumps A to F are indicated. Bottom: JKT B-band image from \protect \cite{perez-gonzalez00}. North is up and East is left.}
\label{figure:clumps}     
\end{figure}

\begin{figure}
\includegraphics[height=9.cm]{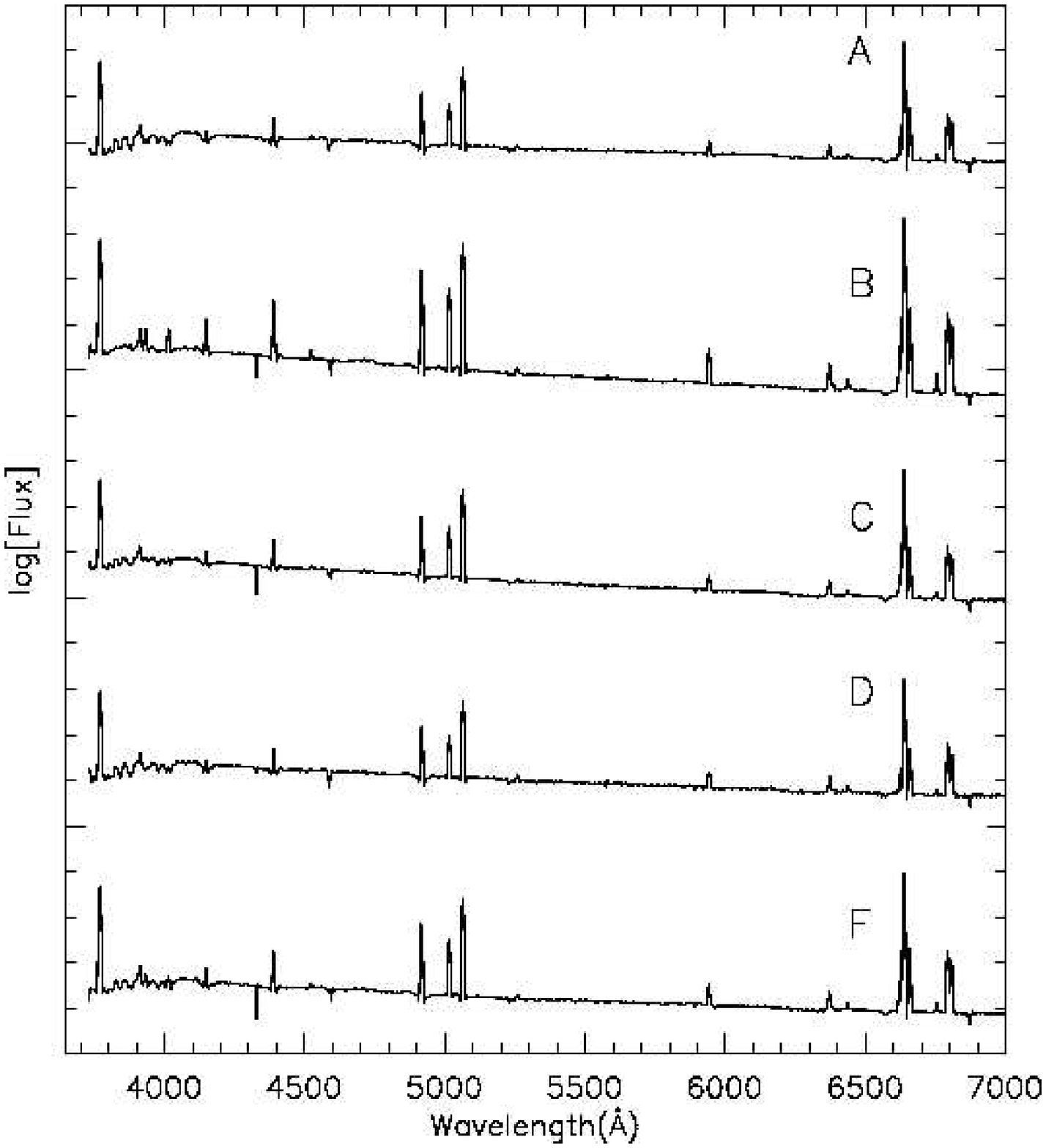}
\includegraphics[height=5.cm]{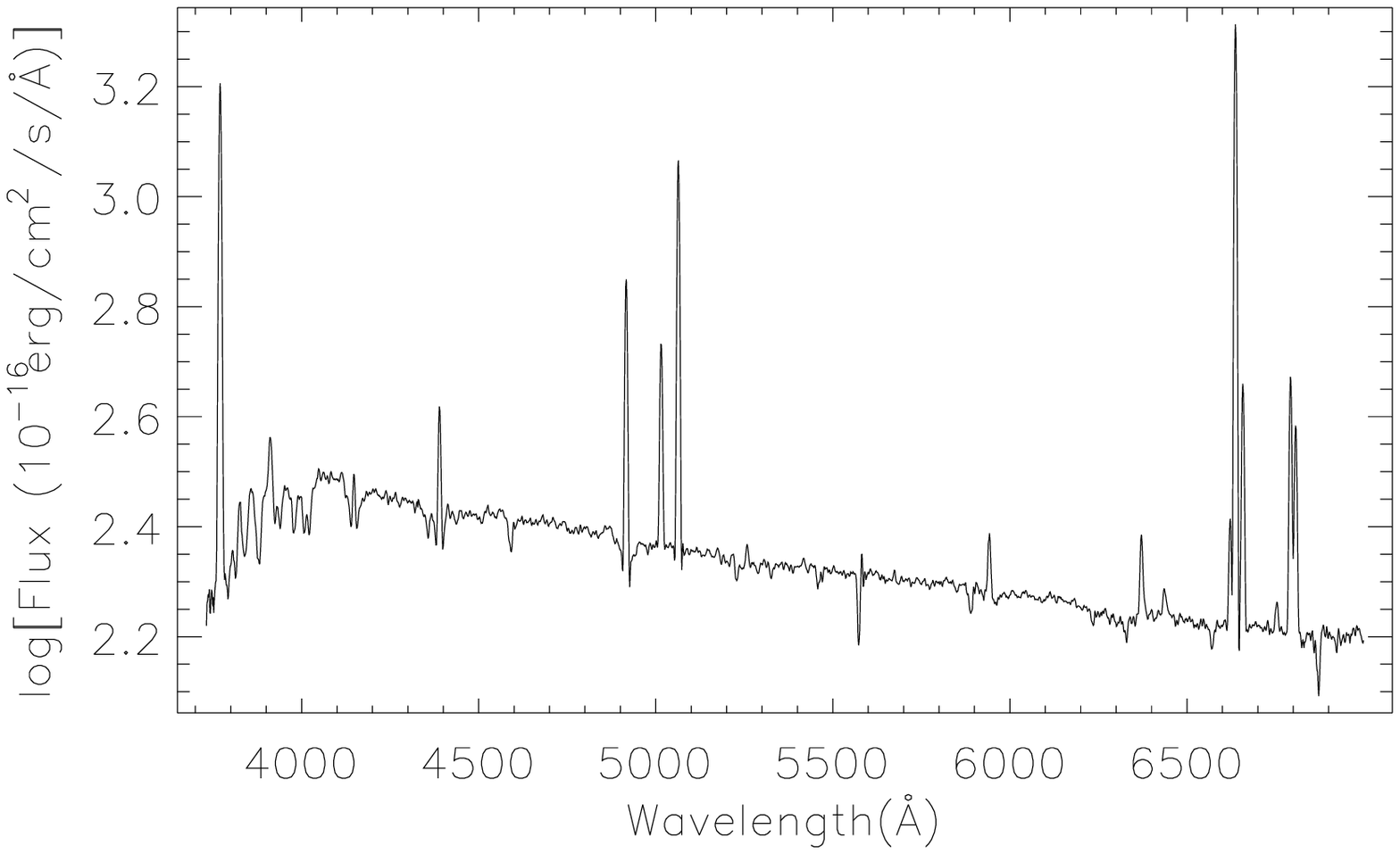}
\caption{Upper panel: Spectra (in logarithmic scale and arbitrarily scaled for better visualization) of different clumps in NGC 7673 obtained summing up the spectra from some pixels defining these regions. Bottom panel: Integrated spectrum (in logarithmic scale) of NGC 7673 obtained co-adding all individual spectra of PPAK dataset over the full field-of-view. All the spectra shown here are obtained with the V300 grating configuration providing a nominal spectral resolution of 10.7 {\AA} FWHM.}
\label{figure:sp}
\end{figure}

For each spectrum, different quantities such as emission line flux, continuum flux and equivalent width, were measured by fitting single Gaussian functions to the observed emission-line profiles using our own software. A two-component (emission and absorption) gaussian fit was performed in order to correct the Balmer lines emission for underlying stellar absorption (see Section~\ref{section:abs}).

Here we analyze in detail the high resolution images obtained with the Hubble Space Telescope in the optical (WFPC2, F555W) and the NOT Telescope in $\rm H\alpha$ (see Figure~\ref{figure:hst_not}) in order to match the spatial position of each clump with the starburst identified in the $\rm H\alpha$ image.
Clump A is composed by several distinct $\rm H\alpha$ knots, with three of them located in the central structure of the bar shape region. In the optical high resolution HST image two bright regions can be distinguished in this area.
Clump B shows a strong $\rm H\alpha$ emission located in one bright knot. The HST image shows clump B to be clumpy with a central peak of emission surrounded by two bright shells. It seems to be situated in a spiral arm connecting this region with the central structure. Clump C  is another strong $\rm H\alpha$ emission region which shows several knots in the optical HST image. In Clump D the $\rm H\alpha$ emission is not seen at the centre of the clump but at the edges. Clump E defined in~\cite{duflot82} can be seen in the optical image as a region of faint emission connected to the central structure. Nevertheless this area is not associated with any $\rm H\alpha$ emission region in particular. At this location (SE of the galaxy) there are several knots of $\rm H\alpha$ emission. And finally clump F is an extended star-formation region composed of many clusters and dominated by a single object~\citep{homeier02} embedded in a strong $\rm H\alpha$ emitting region.

The spatial location of each clump in this paper is selected using the B-band and $\rm H\alpha$ PPAK maps (see Figure~\ref{figure:clumps}). The spectrum
corresponding to each clump is computed by adding up the spectrum from
several pixels (see Figure~\ref{figure:sp}). The integrated spectrum of the galaxy is computed co-adding the spectra of all pixels in PPAK dataset over the full field-of-view. 

The properties computed for different clumps in the following sections are derived from the analysis of the spectrum corresponding to each clump. Moreover,  we have checked the agreement with the same properties derived from the mean values obtained from the computed map in the regions where the clumps are located.

\subsection{Stellar absorption correction}
\label{section:abs}

To obtain an accurate value of the fluxes of the Balmer emission lines we
take into account the presence of an underlying stellar
absorption. The stellar absorption can considerably affect the calculation of quantities such as the extinction.
To solve this problem, two Gaussian functions are fitted
in those cases where the absorption wings are visible (for example, that is the case for the $\rm H\beta$ Balmer line).
For this purpose we use the V1200 configuration data cube whose spectral resolution ($\rm FWHM \approx
2.5 \AA$) allows us to better fit the absorption wings in the $\rm H\beta$
Balmer line. 

For each spectrum with enough signal-to-noise (SNR larger than 15 in the
continuum), two different fits are performed. First, the $\rm H\beta$
Balmer line is fitted only with a Gaussian model in emission. Second, the same data is fitted with two Gaussian models, one in
emission and another in absorption. We tied the centroids of the
gaussian in emission and absorption to be the same. The gaussian
absorption width is upper limited to the value found when the total
galaxy spectrum is fitted and pixels located at the wings of the $\rm H\beta$ absorption are double weighted. The best fit is based in a $\chi^{2}$
scheme, and we accept the two gaussian fit as the best fit when the
reduced $\chi^{2}$ improves at least a $15\%$ with respect to the
single gaussian fit. 
All the spectra in our data cube are fitted with
two gaussian except the spectra corresponding to clump B, where we
fit only a single gaussian in emission and then the $\rm H\beta$ absorption equivalent width is considered to be zero or negligible.

The goodness of our fitting method is checked using simulated
spectra. With signal-to-noise larger than 15 in the continuum, this
method is able to measure absorption equivalent widths as small as
$\rm \approx 2.5 \AA$ with relative uncertainties smaller than 15\%. For larger absorption equivalent
widths the uncertainty decreases: for $\rm EW_{abs} > 4 \AA$ we find a less than 10\% relative error.

In Figure~\ref{figure:ewabs} the $\rm H\beta$ absorption equivalent width ($\rm EW_{abs}$) map and some representative fits in different regions are shown. 
The $\rm EW_{abs}$ values are in the range 0 - 6.6 \AA. 

\cite{duflot82} employed a crude method to find $\rm EW_{abs} \approx 8~\AA$ for the $\rm H\beta$ Balmer line in clump A and find no evidence of absorption wings in clumps B, C, and D.
In Table~\ref{table:data} we summarize the $\rm H\beta$ absorption equivalent widths found for different clumps and the integrated galaxy spectrum. These values are measured fitting the spectra corresponding to each clump.
While in clump B the underlying Balmer absorption is clearly less important, in other regions the stellar absorption is more significant, with computed values of $\rm EW_{abs} \approx 5.6~\AA$ in clump A, or 6.5 AA in regions between clumps A and D. When the integrated galaxy spectrum is fitted the $\rm H\beta$ absorption equivalent width is $\rm 4.8\pm0.5 \AA$. 

\begin{figure*}
   \centering
   \includegraphics[angle=0,width=18.cm, clip=true]{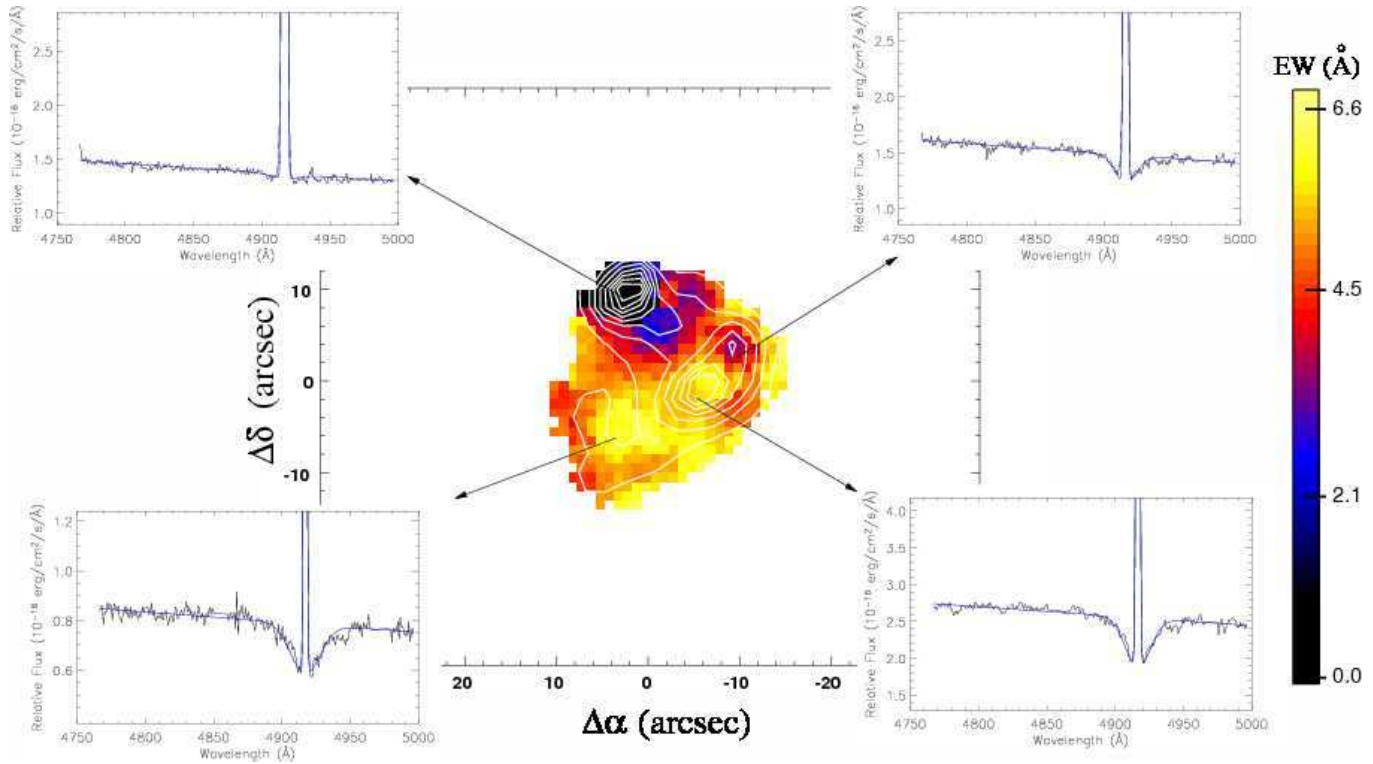}
   \caption{$\rm H\beta$ stellar absorption equivalent width map
   computed from the double gaussian fit method described in
   Section~\ref{section:abs}. Contours from $\rm H\alpha$ emission are
   over-plotted in white colour. Fitted spectra from different regions
   in NGC 7673 are shown.}
\label{figure:ewabs}
\end{figure*}

\subsection{Emission Line Fluxes}

In this paper we focus on the study of the physical properties of the gas through the analysis of the observed emission lines.
In the V300 setup configuration the main emission lines observed are:  [OII]$\lambda3727$, $\rm H\beta$,
[OIII]$\lambda\lambda$4959,5007, [HeI]$\lambda5876$, [OI]$\lambda6300$, $\rm H\alpha$, [NII]$\lambda\lambda$6548,6584, [SII]$\lambda\lambda$6717,6731. The wide wavelength range of this configuration allows us to derive not only line emission fluxes maps but line emission ratios in a consistent way. 

The emission lines in each spectrum of the data cube were fitted by single gaussian profiles using our own software tools. We calculate the error in the line fluxes from the quadratic sum of the error coming from the absolute flux calibration and the error from the expression $\rm \sigma_{l}=\sigma_{c}N^{1/2}[1+EW/N\Delta]^{1/2}$ \citep{castellanos00} where $\sigma_{c}$ is the standard deviation in a box centred close to the measured emission line; N is the number of pixels used in the measurement of the line flux; EW is the equivalent width of the line and $\Delta$ is the wavelength dispersion in $\rm \AA / pixel$. 
Only the spectra with a minimum S/N of 15 in the flux detection are considered to ensure a relative uncertainty less than 10\% in the flux determination.

Some of the emission lines are fitted simultaneously ($\rm H\alpha+[NII]\lambda\lambda6548,6584$, [OIII]$\lambda\lambda4959,5007$, [SII]$\lambda\lambda6717,6731$). In this case, the best fit is found by forcing all the lines fitted simultaneously to have the same velocity shift and width. The continuum under the emission line was fitted at the same time using a linear or quadratic function.

We perform a pixel to pixel correction to the $\rm H\beta$ and $\rm H\alpha$ emission
line fluxes computed from the V300 data cube, using the stellar
absorption EW computed from $\rm H\beta$ Balmer line gaussian fits in the V1200 data cube as explained in section~\ref{section:abs}. We assume the stellar absorption
correction in $\rm H\beta$ and $\rm H\alpha$ are the same (\cite{kurucz92} established that the $\rm H\alpha$ and $\rm H\beta$ equivalent widths are equal within a 30 per cent uncertainty). 
In the case of $\rm H\alpha$ emission, the stellar
absorption corrected fluxes are not significantly different from the
uncorrected fluxes (relative differences are in the range 0-15\% with 5\% relative difference for the integrated galaxy flux). 
For $\rm H\beta$ emission fluxes the differences are much larger with relative differences over 30\% for the integrated galaxy flux and the majority of pixels except for clump B (0\%) and clump F (15\%).
To estimate the error in the $\rm H\alpha$ and $\rm H\beta$ absorption corrected fluxes we take into account the uncertainties from the absorption equivalent width and continuum flux. The main source of error is coming from the uncertainty in the absolute flux calibration.

\subsection{Maps generation}

Finally we generate maps of spectral features such as emission line fluxes, stellar absorption equivalent widths, extinction and other properties from the $regularized$ flux-calibrated data cube using our own software tools. Once we have a collection of values with the information of the measured spectral feature in each pixel we can represent them using the relative position of each pixel in the sky. The computed maps consist of a 72 $\times$ 63 pixels grid where the spatial resolution element (fibre size of 2.7'') is sampled by approximately 3 $\times$ 3 pixels. We decide to work with the spectral information in each pixel instead of in each fibre due to the application of DAR correction as it is explained in Section~\ref{section:datareduction}.

To be able to compare our generated PPAK maps with other images we
perform an astrometric calibration of our PPAK images using the $\rm
H\alpha$ emission PPAK map and the $\rm H\alpha$ narrow-band image from NOT~\citep{perez-gonzalez03} with known astrometry. The NOT image is convolved
with a gaussian kernel function to match the PPAK spatial
resolution. The positions of different HII regions in both images PPAK
and NOT are used to compute an astrometric
solution using several tasks in IRAF package. 
The astrometric calibration is done by forcing the same $x$ and
$y$ scale (1 arcsec/pixel) with an error given by a $rms$ of 0.09 arcsec
and 0.14 arcsec in the $x$ and $y$ axes respectively.

\begin{table*}
\centering
\caption{Summary of PPAK results for NGC 7673 and the main clumps. Col. 1: Region from which the spectrum is analysed. Col. 2 : Measured stellar absorption equivalent width for $\rm H\beta$ Balmer line in {\AA}. Col. 3: Measured emission equivalent width for $\rm H\alpha$ Balmer line with typical errors of 0.5{\AA}. Col. 4: Colour excess obtained from the $\rm F(H\alpha)/F(H\beta)$ ratio. Col. 5: Extinction corrected $\rm H\beta$ luminosity in units of $\rm 10^{41} erg/s$. Col. 6: Extinction corrected $\rm H\alpha$ luminosity in units of $\rm 10^{41} erg/s$ . Col 7: Star formation rate in $\rm M_{\odot}/yr$ computed from extinction corrected $\rm H\alpha$ flux. Col. 8: Star formation rate surface density in $\rm M_{\odot}/yr/kpc^{2}$.  Col 9, 10 and 11: Emission line ratios computed from the extinction corrected fluxes.}
\vspace{0.5cm}
\begin{tabular}{l c c c c c c c  c c  c }
\hline
Region & $\rm EW_{abs}$ &$\rm EW(H\alpha)$ & E(B-V) & $\rm L(H\beta)$ & $\rm L(H\alpha)$ & SFR  & $\Sigma_{\rm SFR}$ &\scriptsize{$\frac{\rm [NII]\lambda6584}{\rm H\alpha}$}& \scriptsize{$\frac {\rm [OIII]\lambda5007}{\rm H\beta}$} & \scriptsize{$\frac{\rm [SII]\lambda\lambda6716,6731}{\rm H\alpha}$}\\  
\scriptsize{ (1) }&\scriptsize{ (2)} &\scriptsize{ (3)} &\scriptsize{ (4)} &\scriptsize{ (5) }&\scriptsize{ (6) }& \scriptsize{ (7) } &\scriptsize{ (8)} & \scriptsize{ (9)} &\scriptsize{ (10)} &\scriptsize{ (11)}\\

\hline
\hline

NGC 7673        	& 4.8$\pm 0.5$ & 106& 0.17$\pm$0.03         &2.7$\pm$0.4         & 7.8$\pm$0.9&6.2$\pm$0.8        & 0.08 &0.15 &1.6 &0.27 	 \\
Clump A  	        & 5.6$\pm 0.6$ & 104& 0.19$\pm$0.03	 &0.36$\pm$0.05     &1.0$\pm$0.1&0.8$\pm$0.1       &0.49 &0.17 &1.6 &0.24      \\
Clump B  	                & 0.0                    & 427& 0.21$\pm$0.01 	 &0.52$\pm$0.06     & 1.5$\pm$0.1&1.2$\pm$0.1      & 0.52 &0.12 &1.9 &0.18      \\
Clump C  	        & 3.1$\pm 0.5$ & 111& 0.16$\pm$0.03 	 &0.11$\pm$0.02     & 0.33$\pm$0.04&0.26$\pm$0.03  & 0.23&0.13 &1.7 &0.24 	 \\
Clump D  	        & 5.1$\pm 0.5$ & 111& 0.12$\pm$0.03 	 &0.034$\pm$0.005 &0.10$\pm$0.01&0.08$\pm$0.01   & 0.09 &0.14 &1.7 &0.27 	 \\
Clump E  	                & 5.2$\pm 0.5$ & 113& 0.20$\pm$0.03 	 &0.032$\pm$0.005 &0.09$\pm$0.01&0.07$\pm$0.01   & 0.10&0.17 &1.2 &0.31      \\
Clump F  	                & 4.1$\pm 0.4$ & 171& 0.20$\pm$0.02	 &0.17$\pm$0.02     & 0.48$\pm$0.05&0.38$\pm$0.04  & 0.38 &0.14 &1.7 &0.24    \\
\hline
\end{tabular}
\label{table:data}	
\end{table*}

%--------------------------------------------------------
\section{Results} 
%--------------------------------------------------------

\subsection{Extinction map}

From the purpose of correcting for dust extinction the observed emission
lines, we have derived the total colour excess ($\rm E(B-V)_{t}$, i.e. the sum of the intrinsic and foreground Galactic reddening) from the
Balmer line ratio $\rm H\alpha / \rm H\beta$. The departure from the
measured $\rm H\alpha / \rm H\beta$ ratio of the theoretically
expected value is produced by the extinction of light due to a screen
of homogeneously distributed interstellar dust.
\begin{equation}
E(B-V)_{t}=K_{\alpha\beta} \;\rm log{\left[ \frac {\rm H\alpha /\rm
H\beta}{2.86} \right]}
\end{equation}
where $\rm K_{\alpha\beta}=[-0.4[k(\rm H\alpha)-k(\rm H\beta)]]^{-1}$;
$\rm H\alpha / \rm H\beta$ is the ratio of the observed $\rm H\alpha$
and $\rm H\beta$ emission lines fluxes;
and 2.86 represents the expected $\rm H\alpha / \rm H\beta$ value for
case B recombination at $\rm T=10^{4} K$ and electron density of 100
$\rm cm^{-3}$~\citep{osterbrock89}. $\rm K_{\alpha\beta}$ depends on the
selected extinction curve and takes the value 2.328 in~\cite{cardelli89}.

\begin{figure}
   \includegraphics[angle=0,width=8.5cm, clip=true]{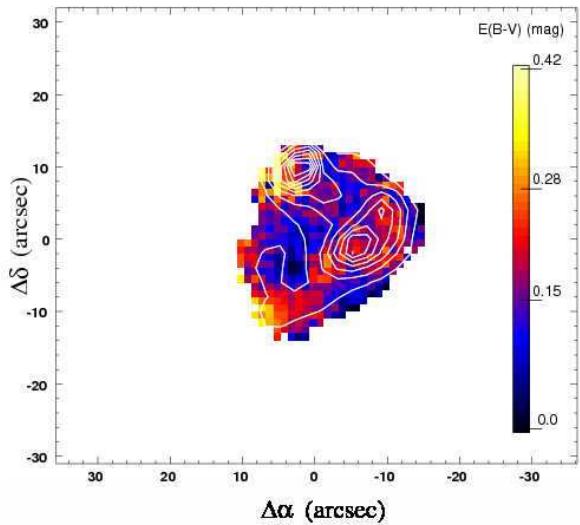}
\hspace{0.5cm}
   \caption{PPAK total colour excess map computed from the Balmer emission lines ratio $\rm F(H\alpha)/F(H\beta)$ using Cardelli extinction law. $\rm H\alpha$ emission is shown in white contours.}
\label{figure:excess}
\end{figure}

The $\rm H\alpha$ and $\rm H\beta$ stellar absorption corrected fluxes are used to derive the colour
excess map (see Figure~\ref{figure:excess}). 
The foreground Galactic reddening at the position of NGC 7673 is $\rm
E(B-V)^{GAL}=0.04$~mag~\citep{schlegel98}. The PPAK colour excess map shows values in the range 0.04 - 0.35 mag in the majority of pixels. 
The dust extinction distribution seems to be related with the position of the main clumps of star formation (except for clump D) showing small spatial variations. In clumps A, B, E, and F the total colour excess is $\rm E(B-V)_{t}\approx 0.20\pm 0.03$ while for clumps C and D is 0.16 and 0.12 respectively. In clump B we find relatively higher values for the colour excess ($\rm E(B-V)_{t}\approx 0.4$) in a region displaced with respect to the maximum $\rm H\alpha$ emission to the east direction. 

These values are in agreement with the colour excess determined by~\cite{pasquali08} using cluster colours in different HST/WFPC2 bands, where the intrinsic cluster reddening $\rm E(B-V)_{i}$ found is lower than 0.4 mag
with the large majority of clusters in this galaxy having $\rm 0. \le
E(B-V)_{i} \le 0.25$ mag. 
They estimate cluster ages based on WFPC2 images and classified the clusters in three samples: young sample with ages $<$ 2.5 Myr, intermediate one with ages in the ragne 2.5 - 8 Myr and the old sample with ages $>$ 8 Myr.
Despite of the large age uncertainty, they point out that the stars clusters in NGC 7673 seem to get younger as their distance in the north-east direction from clump A increases.

From the Balmer decrement we determine for the integrated galaxy spectrum $\rm E(B-V)_{t}=0.17$ mag. The extinction $\rm A_{\rm H\alpha}$
computed from this colour excess is 0.40 mag ($\rm A_{\rm V}=0.49$) using the extinction curve from~\cite{cardelli89}.

The UV wavelength can provide us an independent attenuation estimation.
We estimate the attenuation in the FUV and NUV bands using the ratio
between the total dust emission in the range 1$-$1000$\rm \mu m$ (TIR)
and the observed stellar emission in the FUV and NUV bands. The TIR
luminosity for this galaxy was derived by~\cite{calzetti00}
from IRAS and ISO data. The FUV and NUV luminosities were taken from
~\cite{gildepaz07}. 
By applying the calibrations
of~\cite{cortese08} 
we find the attenuation in the FUV and NUV to be $A_{\mathrm{FUV}}=1.08$\,mag and
$A_{\mathrm{NUV}}=0.86$\,mag, respectively. The slope of the UV
spectrum or, equivalently, the FUV$-$NUV colour, is known to be tightly
correlated with the UV extinction in starburst galaxies~\citep[see e.g][]{calzetti94,meurer99}. Considering
that NGC~7673 has $\mathrm{FUV}-\mathrm{NUV}=0.25$, our values of
$A_{\mathrm{FUV}}$ and $A_{\mathrm{NUV}}$ are fully
consistent with what is expected for a starburst galaxy with that UV
colour.

We can use the previous values of the attenuation in the UV to
estimate the attenuation in the V band. While in both cases we are
referring to the extinction in the stellar continuum, stars
contributing to the optical emission are usually less embedded
within the dust than those dominating the UV emission. To take this fact
into account, we combine the MW extinction curve of~\cite{cardelli89} with the simple sandwich model for dust geometry described in~\cite{boselli03}, resulting in $A_{\mathrm{V}}=0.35$\,mag. 

The extinction computed from the Balmer decrement is $\rm A_{\rm V}=0.49$ mag. This value corresponds to the extinction for the gas, it has to be corrected with the 0.44 factor found by~\cite{calzetti94} to be compared with the extinction derived from the UV colours.
After this correction, the stellar extinction results 0.13 mag lower compared with the attenuation in the V band using UV colours. Anyway, the 0.44 factor is found for most dusty starbursts and this is not the case for NGC 7673 as it is shown from the low extinction values computed in this section.

\begin{figure}
\centering
   \includegraphics[angle=0,width=8.5cm, clip=true]{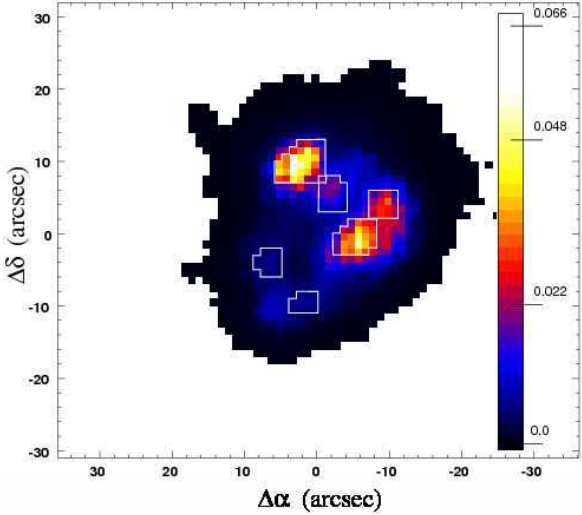}
   \caption{Star formation rate density map in $\rm M_{\odot}/yr/arcsec^{2}$ units, computed from reddening-corrected $\rm H_{\alpha}$ flux PPAK map. SFR are computed in regions enclosed by white colour.}
\label{figure:sfr}
\end{figure}

\subsection{Star formation rates}

Different SFR measures based on $\rm H\alpha$ emission can be found in the literature for this galaxy: 
\cite{schmitt06} derived a lower limit for the SFR of $\rm 1.3 M_{\odot}/yr$
for an observed $\rm H\alpha$ flux that was corrected for Galactic extinction but not for intrinsic reddening. In~\cite{pasquali08}, the observed $\rm H\alpha$ flux from~\cite{mcquade95} is de-reddened using E(B-V)=0.54 mag (derived from the Balmer decrement using $\rm H\alpha$ and $\rm H\beta$ fluxes in~\cite{mcquade95}) and E(B-V)=0.36 mag (estimated for the exponentially decaying star formation history). They finally obtained a SFR between 3.3 and 5.2 $\rm M_{\odot}/yr$.
\cite{pisano01} derived a value of SFR=$\rm 23.5 M_{\odot}/yr$ based on~\cite{gallego96} long-slit spectroscopic data. After reviewing how this SFR value is obtained we find the $\rm L(H\alpha)$ published in~\cite{gallego96} is highly overestimated due to the aperture correction applied. The same authors revisited this object in \cite{perez-gonzalez03} and the SFR derived from the $\rm H\alpha$ imaging study is SFR=9 $\rm M_{\odot}/yr$.
Furthermore, NGC 7673 is also a well-known luminous FIR source~\citep{sanders96}.~\cite{garland05} derived a IR-SFR of 5.5 $\rm M_{\odot}/yr$ (IRAS data) in agreement with the derived SFR from the 1.4 GHz radio continuum in NGC 7673 using the prescription of \cite{bell03}.

From the integrated galaxy spectrum we compute a total extinction-corrected luminosity $\rm L(H\alpha)=(7.8\pm1.0)\times 10^{41} erg s^{-1}$, from which a~$\rm SFR=(6.2\pm0.8) M_{\odot} yr^{-1}$ is derived (using the relation from ~\cite{kennicutt98}). The $\rm H\alpha$ luminosity derived by~\cite{perez-gonzalez03} ($\rm L(H\alpha)=11.39\times10^{41}erg s^{-1}$) is higher compared with our value mainly due to the absolute flux calibration (see Section~\ref{section:datareduction}) and the extinction correction applied. They used the ratio $\rm F(H\alpha)/F(H\beta)=4.21$ to correct the observed fluxes, and we find for the integrated galaxy spectrum the ratio $\rm F(H\alpha)/F(H\beta)= 3.4$.
Our $\rm H\alpha$-based SFR estimation is in agreement with the SFR derived from infrared and radio continuum fluxes~\citep[][SFR = 5.5 $\rm M_{\odot}yr^{-1}$]{garland05}.

The SFR can also be estimated from the $\rm [OII]\lambda3727$ luminosity. We use the~\cite{kennicutt98} relation ($\rm SFR(M_{\odot}yr^{-1})=(1.4 \pm 0.4)\times10^{-41}L([OII])(erg s^{-1})$), where the observed [OII] luminosity must be corrected for extinction; in this case the extinction at $\rm H\alpha$ because of the manner in which the [OII] fluxes were calibrated to obtain that relation. We compute a total [OII]-based SFR of 8.5 $\rm M_{\odot}yr^{-1}$, resulting in the range 6.1-11.0 $\rm M_{\odot}yr^{-1}$ when we take into account the error associated in the relation L([OII])-SFR. 

We derive a SFR surface density map based on $\rm H\alpha$ extinction-corrected PPAK fluxes (see Figure~\ref{figure:sfr}) using the relation in~\cite{kennicutt98}. 
The star formation is located mainly in clumps A, B, C and F as can be seen in Figure~\ref{figure:sfr}. Clump B shows the highest $\rm H\alpha$ luminosity with a SFR surface density of 0.52 $\rm M_{\odot}yr^{-1}kpc^{-2}$. 
Clumps D and E are regions with low $\rm H\alpha$ luminosity and their contribution to the total star formation rate is not significant.
Although clump E was selected in this work for comparison reasons with the results by~\cite{duflot82}, this region is not associated with any star-forming region in particular, showing a low SFR.
More than 50$\%$ of the total star formation rate in this galaxy is not located in the defined clumps.
The star formation rates derived from the integrated galaxy spectrum and the spectra corresponding to the different clumps marked in Figure~\ref{figure:sfr} are tabulated in Table~\ref{table:data}.

\begin{figure}
\centering
\includegraphics[height=5cm]{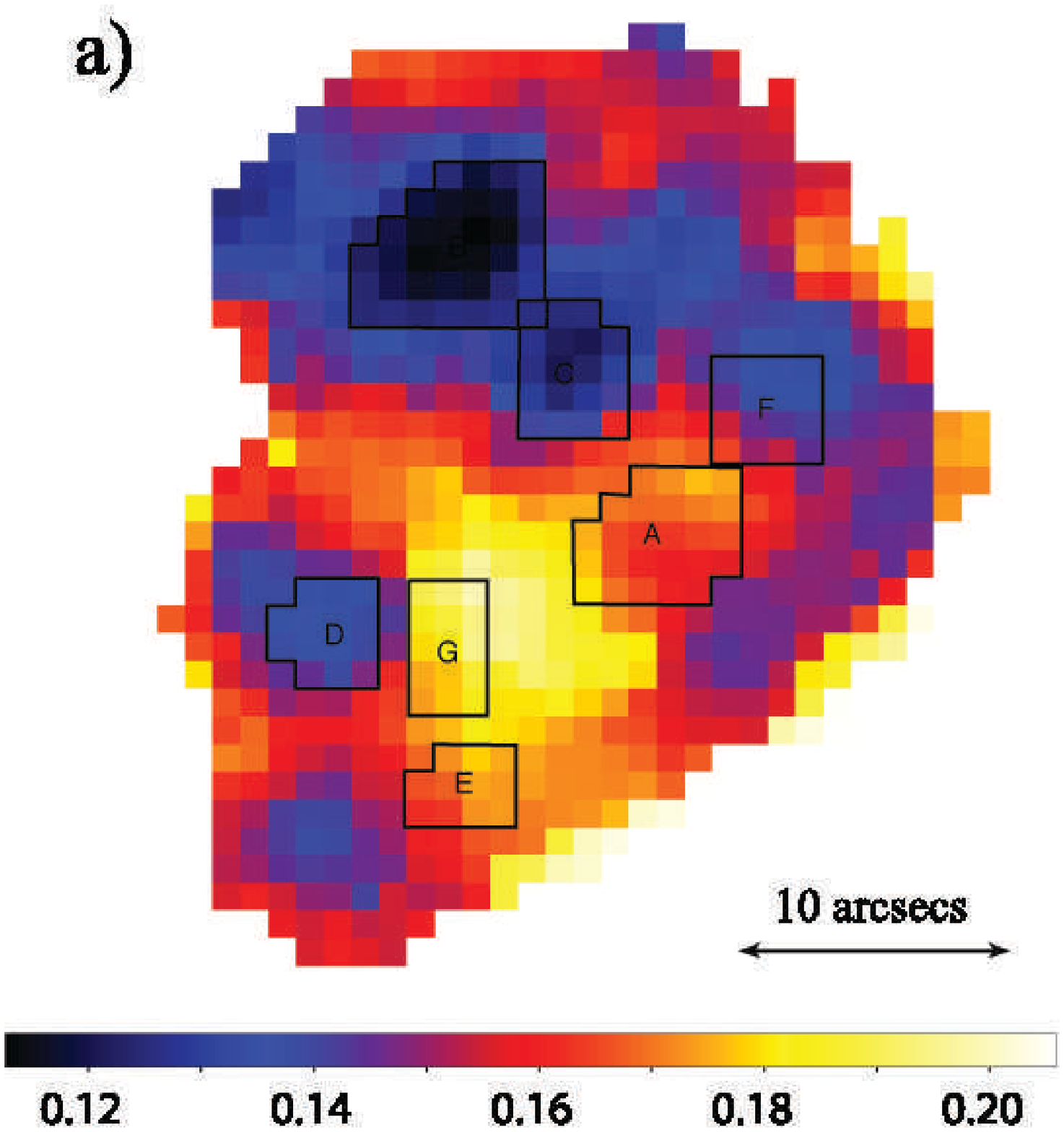}
\hspace{1.cm}
\includegraphics[height=5cm]{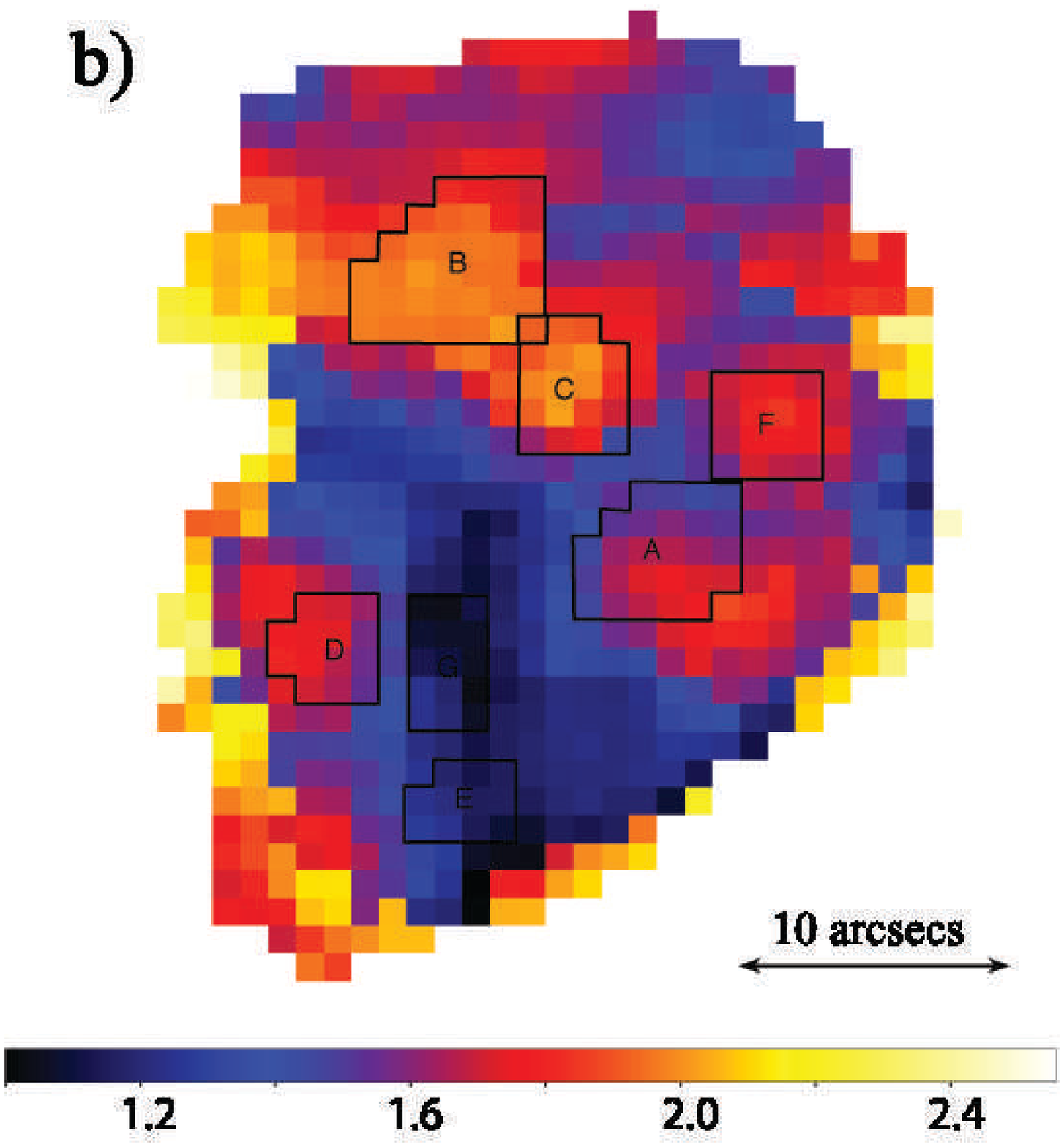}
\hspace{1.cm}
\includegraphics[height=5cm]{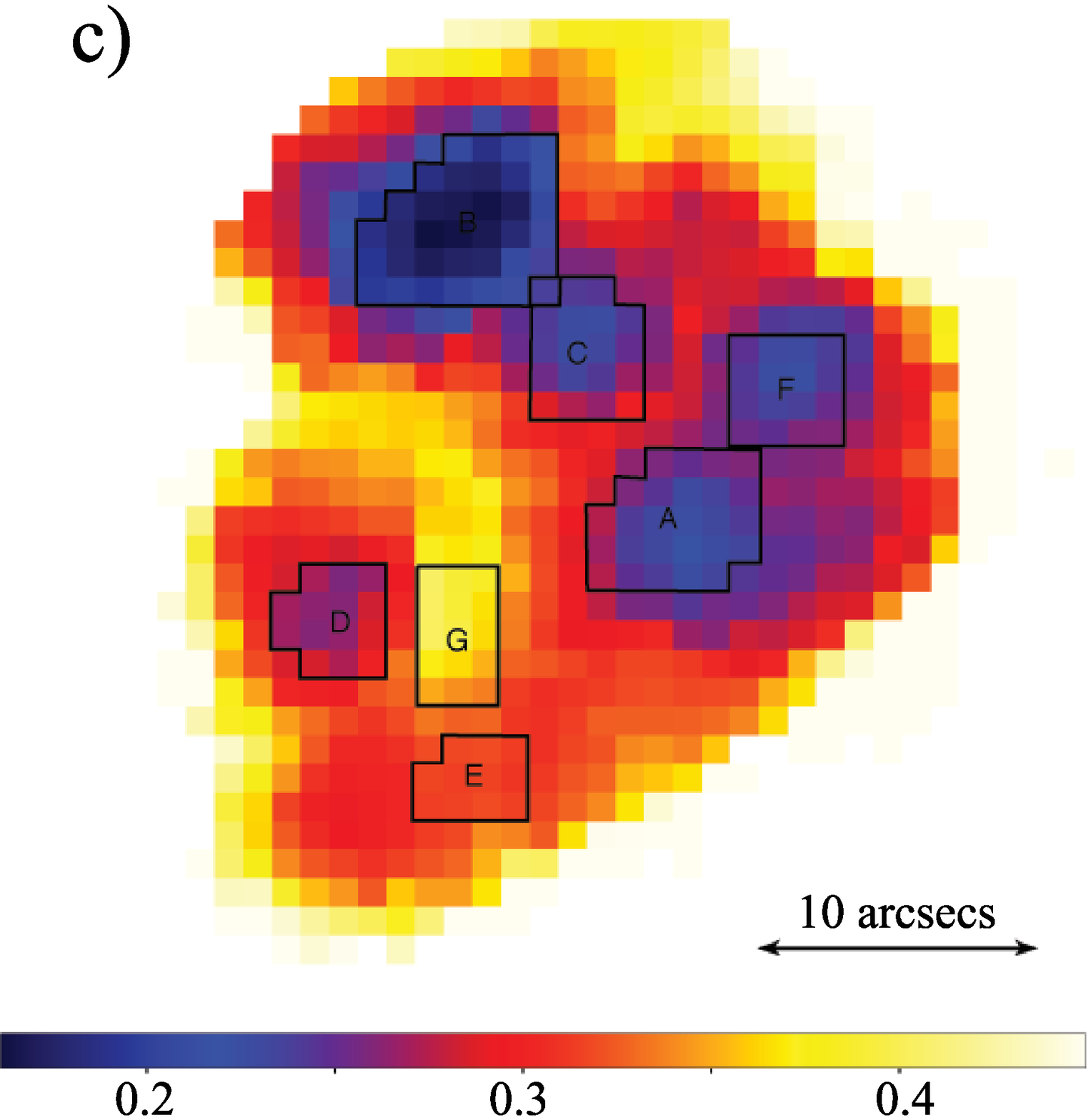}
\caption{PPAK emission line ratio maps: a) $\rm [NII]\lambda6584/\rm H\alpha$, b) $\rm [OIII]\lambda5007/\rm H\beta$ and c) $\rm [SII]\lambda\lambda6716,6731/\rm H\alpha$. The position of the main clumps are overlaid.}
\label{figure:mapratios}
\end{figure}

\begin{figure}
\centering
\includegraphics[height=9.cm]{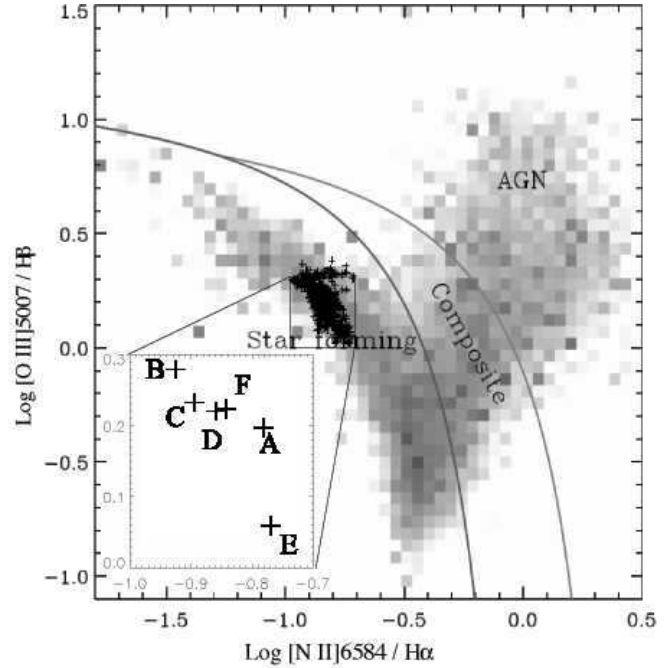}
\caption{Diagnostic diagram $\rm [NII]\lambda6584/H\alpha$ versus $\rm [OIII]\lambda5007/H\beta$ is shown for SDSS galaxies (grey pixels) from \protect \cite{brinchmann04}. The values for every pixel in the PPAK map are included in black crosses.
The location of clumps A-F is also shown in an included figure.}

\label{figure:ratio}     
\end{figure}

\begin{figure}
\centering
\includegraphics[height=6.cm]{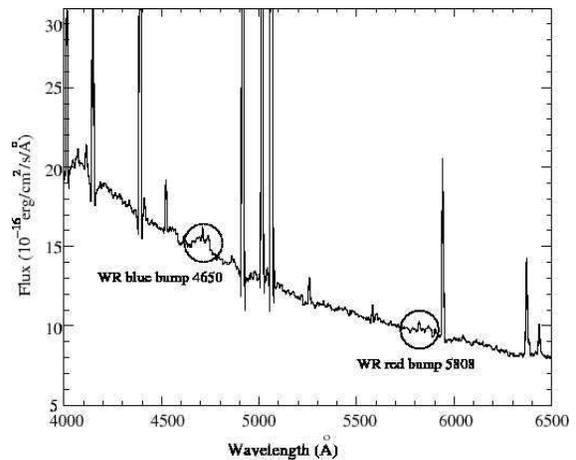}
\caption{Sum of the spectra in clump B where the WR stellar population is detected. The blue and red WR bumps can be observed.}
\label{figure:wr}     
\end{figure}

\subsection{Emission line ratio maps}
\label{ratios}
We have computed different emission line ratios maps (i.e. $\rm
[OIII]\lambda5007/H\beta$, $\rm [NII]\lambda6584/H\alpha$, $\rm
[SII]/H\alpha$) to study the possible presence of an AGN or signs of shocks in this
galaxy {(see Figure~\ref{figure:mapratios}).  The position of the different clumps except for clump E are very well traced in the $\rm [SII]/H\alpha$ map with clumps corresponding to regions of low emission line ratio (0.24).  Clump B in this map shows the lowest value (0.18) and clump D the highest one (0.27).  Region G is defined as a new region between clumps A and D. In this region the presence of the underlying stellar population is most noticeable (see Figure~\ref{figure:ewabs} of absorption stellar equivalent width) and the $\rm [SII]/H\alpha$ ratio reaches the value of 0.40. The map corresponding to the ratio $\rm [OIII]\lambda5007/H\beta$  shows higher values, corresponding to higher ionization regions, at the position of clumps with values of 1.6, showing a slightly higher value 1.9 at the position of clump B. The different clumps show values of $\rm [NII]\lambda6584/H\alpha$ very similar in the range 0.12 (for clump B) and 0.17 (for clump A). Only region G shows a higher value of 0.20 for this ratio. \cite{duflot82} found a similar distribution of the $\rm [NII]\lambda\lambda6458,6584/H\alpha$ ratio. In their analysis they computed the highest ratio value in clump A and lowest one in clump B.
In Figure~\ref{figure:ratio} the diagnostic diagram $\rm
[NII]\lambda6584/H\alpha$ versus $\rm [OIII]\lambda5007/H\beta$ is
shown for every pixel in the PPAK map. 
For comparison, the values for
SDSS galaxies are plotted in the same figure. We find values of $\rm [NII]\lambda6584/H\alpha$ ratio in the range 0.1-0.2 and $\rm [OIII]\lambda5007/H\beta$ values in the range 1-2.5.
The location of clumps A-F in this
diagram is shown in the same Figure and situate the clumps in a sequence of metallicity (as suggested by the several sets of theoretical photoionization models described in~\cite{dopita86}) where clumps E\footnote{Actually, clump E, although defined in \cite{duflot82}, is not associated with any $\rm H\alpha$ emission region in particular.} and A would have higher metallicity values, clumps C, D and F intermediate metallicity values, and clump B the lowest one.
These ratio values (see Table~\ref{table:data}) are clearly in agreement with those of HII
regions of intermediate metallicity and with no signs of AGN activity~\citep{osterbrock89}. The $\rm [SII]/H\alpha$ map is also computed showing values $<0.4$ for all the pixels. The location of these values in the diagnostic diagram $\rm [SII]/H\alpha$ versus $\rm [NII]\lambda6584/H\alpha$ is compatible with a starburst galaxy~\citep[see Figure 12 in][]{rickes08} ruling out the presence of LINER or shocks as the ionization source in this galaxy.

\subsection{Wolf-Rayet stellar population}

Wolf-Rayet (WR) stars are evolved, massive stars, which are losing mass rapidly by means of a very strong stellar wind~\citep{maeder94}.
The presence of WR stars can be inferred from the detection of WR bumps around $\lambda4650$ \AA (blue bump) and $\lambda5808$ \AA (red bump), which are generally a blend of HeII and several metal lines.

We detect the presence of WR stellar population in NGC 7673. The blue and red WR bumps can be observed in several fibres at the position of clump B. The sum of the spectra where WR bumps are detected is shown in Figure~\ref{figure:wr}. The WR bumps intensity maximum coincides spatially with the position of $\rm H\alpha$ peak emission in clump B. The detection of a WR stellar population at this clump indicates this is a young (less than 4 Myr) burst of massive stars. In agreement with this detection~\cite{pasquali08} found cluster ages younger than 2.5 Myr and in the range 2.5 - 8 Myr for the majority of star clusters located in clump B. 

\begin{figure*}
\centering
\includegraphics[height=8.cm]{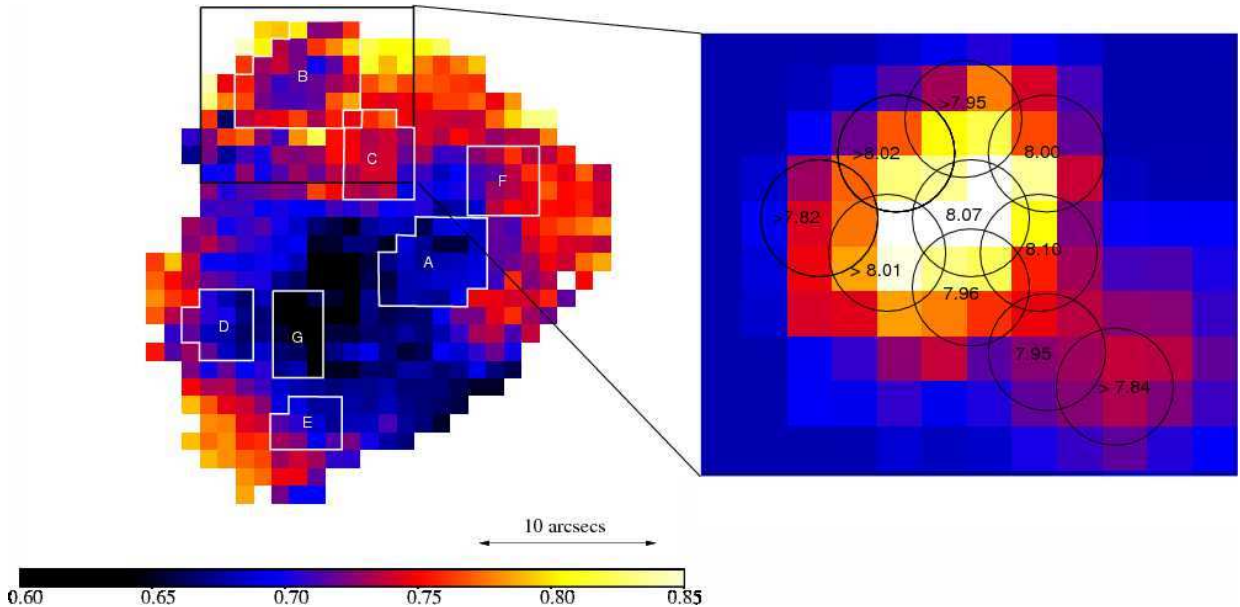}
\caption{Left: $\rm R23=log(([OII]\lambda3727+[OIII]\lambda\lambda4959,5007)/H\beta)$ map with regions where the oxygen metallicity (using R23 index) is computed overlaid. Right: Detail of the $\rm H\alpha$ emission in clumps B and C. The position of some fibres with the computed oxygen metallicity is overlaid. Oxygen metallicity (12+log(O/H)) is derived in five fibres located at clump B and C where the emission line $\rm [OIII]\lambda4363$ is detected (5-sigma level). In some fibres we compute a metallicity lower limit.}
\label{figure:abund}     
\end{figure*}

\begin{table}
\centering
\caption{Metallicity estimations for each region derived from the R23 index. Col. 1: name of each region. Col. 2: $\rm R23=log(([OII]\lambda3727+[OIII]\lambda\lambda4959,5007)/H\beta)$ values with typical error of 0.07 dex. Col. 3: Derived oxygen metallicity using R23 method described in \protect \cite{kewley02} for the low metallicity region, taking into account the ionization parameter. Col. 4: Measured $\rm M_{B}$ (corrected for Galactic extinction) in magnitudes with typical error of 0.1 mag.}
\vspace{0.5cm}
\begin{tabular}{l c  c c }
\hline
Region & R23 & 12+log(O/H) & $\rm M_{B}$  \\  
\hline
\hline

NGC 7673        	&0.72 &8.20$\pm$0.15 & -20.2     	 \\
Clump A  	        &0.68 &8.12$\pm$0.12 & -17.9	       \\
Clump B  	                &0.73 &8.23$\pm$0.15 & -17.3         \\
B Surroundings         &0.76 &8.30$\pm$0.20    & -       \\
Clump C  	        &0.73 &8.21$\pm$0.16 & -16.9 	 \\
Clump D  	        &0.69 &8.14$\pm$0.12 & -15.6		 \\
Clump E  	                &0.69 &8.15$\pm$0.14 & -15.3    \\
Clump F  	                &0.73 &8.21$\pm$0.16 & -16.7      \\
Region G                  &0.64 &8.07$\pm$0.11 &   -       \\
\hline
\end{tabular}
\label{table:r23}	
\end{table}

\begin{table}
\centering
\caption{Metallicity estimations for each fibre in Figure~\ref{figure:abund} derived from the Te-method and R23 ratio as explained in Section~\ref{section:metal}. Col. 1: fibre number. Col. 2: Oxygen metallicity computed from Te-method. Col. 3: Derived oxygen metallicity using R23 method.}
\vspace{0.5cm}
\begin{tabular}{l c c }
\hline
Fibre & $\rm 12+log(O/H)_{Te}$ & $\rm 12+log(O/H)_{R23}$  \\  
\hline
\hline
1  & 7.96$\pm$0.05  &  8.14$\pm$0.14 \\
2  & 8.00$\pm$0.05  &  8.14$\pm$0.14 \\
3  & $>$ 8.02       &  8.33$\pm$0.25 \\
4  & 8.07$\pm$0.04  &  8.06$\pm$0.11 \\
5  & 7.95$\pm$0.06  &  8.35$\pm$0.27 \\
6  & $>$ 7.82       &  8.31$\pm$0.22 \\
7  & $>$ 8.01       &  8.11$\pm$0.13 \\
8  & $>$ 7.95       &  8.28$\pm$0.19 \\
9  & 8.10$\pm$0.05  &  8.07$\pm$0.12 \\
10 & $>$ 7.84       &  8.12$\pm$0.13 \\

\hline
\end{tabular}
\label{table:fibresabund}	
\end{table}

\subsection{Metallicity}
\label{section:metal}

\cite{duflot82} determined almost a constant oxygen abundance in the clumps studied in NGC 7673 (12+log(O/H)$\approx$8.6) using empirical methods described by~\cite{alloin79} and~\cite{pagel79}.~\cite{storchi94} obtained a similar value of oxygen abundance (12+log(O/H)=8.48) using the calibration diagrams from~\cite{pagel79} to obtain T[OIII].

A precise measurement of the weak auroral forbidden emission line $\rm [OIII]\lambda4363$ is needed to give an accurate determination of oxygen abundance in gaseous ionized nebulae. This emission line which is temperature sensitive correlates with the overall abundance~\citep{osterbrock89}, being relatively strong in very low metallicity systems
but undetectable for even moderately low metallicity galaxies ($\rm 12+log(O/H) > 8.3$)~\citep{denicolo02}. 

We inspected all the spectra in NGC 7673 and for some fibres the detection of the emission line $\rm [OIII]\lambda4363$ is possible with enough signal-to-noise ratio ($5\sigma$-level).
In particular we are able to detect $\rm [OIII]\lambda4363$ in a few spectra corresponding to clump B where the stellar absorption over the $\rm H\gamma$ emission is not so important and the emission flux for $\rm [OIII]\lambda4363$ can be measured reliably. We apply the Te-method to measure gas metallicity, using the routine {\it{abund}} of the {\it{nebular}} package of the STSDAS/IRAF for this task. 
We assume a two-zone model for the star-forming nebula, a medium-temperature zone where oxygen is doubly ionized and a low-temperature zone where oxygen is assumed to be singly ionized and neutral. 
The electron temperature in the medium-temperature zone, $\rm T_{e}(OIII)$ is derived from the [OIII]($\lambda$4959+$\lambda$5007)/$\lambda$4363 line intensity ratio~\citep{izotov94}. The temperature in the low-temperature zone, $\rm T_{e}(OII)$ is estimated using the relation $\rm t_{e}(OII)=2[t_{e}(OIII)^{-1}+0.8]^{-1}$, proposed by~\cite{pagel92} based on HII region models of~\cite{stasinska90}, where $t_{e}$ are temperatures measured in units of $\rm 10^{4} K$.
The electron density ($\rm N_{e}$) is estimated from the [SII]$\lambda$6716/$\lambda$6731 ratio.
The computed total oxygen abundance is the sum of all measured oxygen ionic abundance,
considering negligible the contribution of $\rm O^{3+}$ to the total abundance.

The error estimation for the oxygen abundance comes mainly from the uncertainty in the $\rm [OIII]\lambda4363$ emission flux. We estimate 10-18$\%$ relative error in this emission flux. We compute new electronic temperatures taking into account this uncertainty, and derive new oxygen abundances. Relative errors in temperature range from 4$\%$ to 7$\%$ and result in final metallicity errors of 0.04-0.07 dex. 

In some fibres located at clump B, we use the signal-to-noise ratio of each spectrum to find an upper limit to the $\rm [OIII]\lambda4363$ line strength in the case this auroral line is not detected. With this estimation we can derive an upper limit to the electron temperature and hence a lower limit for the oxygen abundance.

The oxygen abundance computed in clump B is in the range $\rm 12+log(O/H)$=7.96 - 8.10, with a value of $\rm 12+log(O/H)=8.07\pm 0.04$ ($\approx$1/4 solar\footnote{We consider $\rm [12+log(O/H)]_{\odot}$=8.69~\cite{allende01}}) for the fibre at the position of the maximum $\rm H\alpha$ intensity. 
Furthermore, the oxygen abundance derived from the sum of spectra in clump B is $\rm 12+log(O/H)=8.02\pm 0.05$.
We attempted to find an estimation of the metallicity summing up the signal from different fibres around clump B. In this case the $\rm [OIII]\lambda4363$ line strength is difficult to measure due to the position of this emission line near to the $\rm H\gamma$ absorption stellar wing. Nevertheless we estimate a lower limit for the oxygen abundance in the surroundings of clump B of $\rm 12+log(O/H)>7.87$.

A metallicity estimation for the rest of regions in this galaxy is obtained using the R23 index ($\rm R23=log(([OII]\lambda3727+[OIII]\lambda\lambda4959,5007)/H\beta)$. We apply the R23 method described in~\cite{kewley02} for the low metallicity region, taking into account the ionization parameter. From the spectrum corresponding to each clump we derived an initial guess for the metallicity using the [OII]/[OIII] and [NII]/[SII] emission line ratios~\citep{charlot01}. Taking a metallicity estimation for each clump, the first estimation of the ionization parameter is derived from the [OIII]/[OII] ratio. The R23 index is then used to determine the oxygen abundance and if this value is significantly different from the initial metallicity guess, the ionization parameter is again computed with the new metallicity estimation. This process is repeated until there is no significant change in the metallicity value which usually happens after the first iteration. The R23 values and the oxygen abundance estimations for different regions in NGC 7673 are shown in Table~\ref{table:r23}. Figure~\ref{figure:abund} shows the R23 map with values in the range 0.6-0.8. For the integrated galaxy spectrum we derive $\rm R23=0.72\pm0.07$ and $\rm q=2\times10^7 cm s^{-1}$ which gives $\rm 12+log(O/H)=8.20\pm0.15$, lower than previous works.
\cite{duflot82} found oxygen abundance of $\rm 12+log(O/H)=8.45$ for clump A and $\rm 12+log(O/H)=8.60$ for clump B using the empirical method based on the $\rm ([OII]+[OIII])/\rm H\beta$ line intensity ratio and the opposite values when the $\rm [OIII]/[NII]$ ratio is employed. They considered the two empirical methods provide abundances in good agreement taking into account the uncertainty inherent to this kind of analysis.

For comparison we computed the R23 oxygen abundance in those fibres where we detect $\rm [OIII]\lambda4363$ (see Figure~\ref{figure:abund} and Table~\ref{table:fibresabund}). The R23 method overestimates the oxygen abundance in fibres 1,2, and 5 (differences of 0.14-0.40 dex). The two methods give similar results in fibres 4 and 9 where the signal-to-noise ratio in the $\rm [OIII]\lambda4363$ detection is higher suggesting that the R23 method is reliable. For the remaining fibres (3, 6, 7, 8 and 10) where the $\rm [OIII]\lambda4363$ auroral line is not detected we only estimate lower limits for the abundances derived with the Te method.

In conclusion, for clump A which dominates the galaxy morphology we derive a low R23 metallicity estimation (0.27 solar). Clump B seems to be surrounded by a region which shows marginally higher metallicity (0.41 solar) compared with the metallicity derived in clump B (0.35 solar) from the same method R23. Clumps D and E with the same metallicity estimation (0.28 solar) seem to belong to the same spiral arm. And finally clumps C and F situated at equidistant positions respect to clump A, have the same derived metallicities (0.33 solar).

\section{Discussion}

\subsection{Nucleus}
\label{section:nucleus}

The location of the nucleus in NGC 7673 is not well determined.~\cite{homeier99} refers to clump A as the ``nucleus''. If this is true we would expect a specific behaviour in the
kinematic, velocity width maps and diagnostic diagrams. From the analysis of the $\rm H\alpha$ kinematic map~PG10 find the position of the kinematic centre in between clumps C and A. This location corresponds to the position of a maximum found in the gas velocity width map and roughly coincides with the photometric and geometric centres of the galaxy.

This latter finding does not, however, rule out the possibility for clump A to host the nucleus. Indeed, our computed extinction map shows at the position of clump A a concentration of extinction although with not very high values. Probably the nucleus of this galaxy is located in this clump but the optical light does not penetrate enough to reveal the optical counterpart of the nucleus. 

Radio continuum emission in this galaxy might help to locate the position of the nucleus.
\cite{condon90} shows VLA data (A-array at 1.46 GHz) mapped at 3" and 1.5" angular resolution. These radio continuum images show an irregular, diffuse morphology characteristic
of a starburst with the two brightest radio peaks matching up he brightest optical clumps A and B. But there is no radio peak emission at the position of the kinematic centre found in PG10.
Furthermore in their high-resolution radio map there is no evidence of any compact source stronger than S=0.5 mJy or of radio jets powered by a nuclear AGN.

When we represent clump A in several diagnostics diagrams we find no sign of any AGN activity (see Figure~\ref{figure:ratio}). Mid-IR spectra can shed further light on the AGN versus starburst nature of the galaxy. In particular, the emission features due to Polycyclic Aromatic Hydrocarbons (PAHs)
are known to be much weaker in AGNs than in starburst galaxies~\citep[see, e.g$.$,][and references therein]{weedman09}. In the particular case of NGC~7673,
mid-IR spectra from the Infrared Spectrograph~\citep[IRS,][]{houck04} onboard Spitzer~\citep{werner04} are publicy available in the Spitzer archive
(program ID 73, PI: J$.$ R$.$ Houck). We used the tool CUBISM~\citep{smith07} to build and inspect the corresponding data-cube. PAHs features are visible
across the whole galaxy, and are particularly bright in clump A, whose spectrum resembles the starburst templates presented in~\cite{weedman09}. 
Moreover, we do not observe strong high-ionization lines of [Ne V] or [O IV] which, if present, would point towards an AGN as the primary ionizing source. 
Therefore, the mid-IR data supports the starburst nature of this galaxy.

\subsection{The peculiarity of clump B}
\label{section:clumpB}
Different properties measured in clump B makes this region peculiar. This clump with bluer colour ($\rm M_{B}=-17.3$) and medium-high star formation rate density ($\rm 0.5 M_{\odot}/yr/kpc^{2}$) represents a compact decoupled kinematic component studied in detail by~PG10.
The HST image of this galaxy shows B region to be clumpy with a central peak of emission surrounded by two bright shells. 
Its bubble morphology is probably the result of the stellar winds associated with massive stars.
Indeed, we detect the presence of Wolf-Rayet stellar population in several fibres in clump B pointing out the presence of massive stars, where young (less than 2.5 Myr old) and intermediate age clusters (between 2.5 and 8 Myr) are found~\citep{pasquali08}. 

While at the centre of clump B we find extinction values similar to clumps A and F, relatively higher values for the colour excess ($\rm E(B-V)_{t}\approx 0.4$) are found in a
region displaced respect to the maximum $\rm H\alpha$ emission to the east direction.
Our measurements of the equivalent width of the underlying stellar population absorption features (e.g., $\rm H\beta$, $\rm H\gamma$) show a peculiarity 
at the location of clump B.
The strength of these features is noticeable throughout the galaxy but it is almost absent at the location of clump B (see Figure~\ref{figure:ewabs}). 
The strength of the equivalent width of these features account for the age of the underlying population. Strong lines are mainly due to a population of A class 
stars and correspond to systems a few hundred thousand years old. Weaker lines are typical of either younger or older populations. 
Nonetheless, clump B shows signs of differentiation, at least marginally, with respect to its environment from the point of view of both, the gas and the stars.

From our metallicity study clump B is a region with low metallicity ($\approx$ 1/4 solar) surrounded by a region with marginally higher metallicity (see Figure~\ref{figure:abund}). 
Furthermore, a detailed study using the R23 index, shows some fibres at the east side of the central fibre in clump B with higher metallicity ($\approx 1.2$ solar). This region is located at the position where~PG10 find maximum values for the velocity width. 

The location of NGC 7673 in the luminosity-metallicity relation is in agreement with the trend observed for intermediate-z star-forming galaxies from~\cite{salzer09} and from intermediate-z LCBGs from~\cite{hoyos05} (see~PG10). But the properties of clump B, being less luminous and with approximately the same metallicity, positioned this region closer to the dwarf irregulars trend~\citep{richer95} as already seen in~PG10. But this region could also be a nearby giant HII region following the $\rm H\beta$ luminosity-velocity width correlation by \cite{melnick87}. 
Nevertheless, as explained in PG10, clump B is brighter and more massive than any of the nearby giant HII regions within their sample, suggesting that an in-falling dwarf galaxy rather than a giant HII region, might be the reason for the peculiarities found in this clump. 

In summary, our 3D data show clump B is a compact region kinematically decoupled with high extinction and star formation rate density and with luminosity and metallicity properties that suggest a dwarf galaxy or a giant HII region nature for this clump.

\subsection{On the nature of NGC 7673}

The star-formation activity in this LCBG galaxy is located in several clumps as it is shown in the star formation rate surface density map. In general, the origin of the starburst activity could be explained by different mechanisms which include, interactions and mergers~\citep{schweizer87,jog92}, bar instabilities~\citep{shlosman90} and kinematic from SNe and stellar winds ~\citep{heckman90}.
Although~\cite{homeier99} excluded an active interaction of NGC 7673 with its companion NGC 7677 as the starburst trigger, they cannot rule out a past interaction. Another possibility suggested by these authors is the capture of a dwarf companion in a minor merger as the main starburst triggering which would account for the presence of the different bursts and the characteristic morphology in this galaxy.

The HST image of this galaxy\footnote{http://www.spacetelescope.org/images/html/heic0205a.html} shows a clearly distorted morphology with an inner disturbed spiral structure. Clump A dominates the galaxy morphology and probably the nucleus of NGC 7673 is hidden by extinction at this position. The ring-like shape with the absence of $\rm H\alpha$ emission at the centre of clump D (as seen in the high spatial resolution map) shows this is an evolved star-forming region decoupled from the spiral structure where region E is located.
Our kinematic study in PG10 shows an asymmetrical ionized gas velocity map, where a decoupled kinematic component is found at the position of clump B. This region results peculiar not only from the kinematic study but also from different physical properties derived in this paper. 
This region, with the highest $\rm H\alpha$ emission equivalent width and star formation surface density, is composed of young (we detect the presence of WR stellar population)  and intermediate age clusters where the underlying stellar population shows the absence of strong absorption features. We find no evidence for neither AGN activity (see Section \ref{ratios}) nor SNe galactic winds (PG10) in this kinematically decoupled component, and from the metallicity and luminosity derived in this work, clump B is in agreement with being an extremely giant HII region or an in-falling dwarf galaxy (see~PG10).

As stated in~\cite{homeier99}, NGC 7673 and NGC 3310, present morphological and spectroscopic similarities in H II and H I data which would lead us
to think in the same triggering scenario for both starburst galaxies. NGC 3310 has been largely classified as a system which is undergoing a minor merger with
a dwarf companion~\citep[see for example][]{mulder96}. A similar scenario would account for the inner bursts in NGC 7673 and the different morphological features,
such as arcs, in its outer parts. The spectroscopic and kinematic properties derived for clump B in this work and in PG10 would lead us to think that this decoupled kinematic component is the dwarf satellite galaxy in the minor merger hypothesis although this needs to be confirmed with further studies.
While~\cite{homeier99} talk about a past minor merger to allow the outer disk to mostly recover, the similarities with NGC 3310, the position of the decoupled kinematic component in the galaxy disk and the young and intermediate age of its starburst seem to support an on-going interaction. 

\section{Summary}

Physical properties of the star-forming regions in the local Luminous Compact Blue Galaxy (LCBG) NGC 7673 are studied in detail using three dimensional data taken with the PPAK integral field unit at the 3.5-m telescope in the Centro Astron\'omico Hispano Alem\'an. From previous ground-based images six main clumps have been identified in this galaxy being four of the clumps strong $\rm H\alpha$ sources (A,B,C and F). We derive integrated and spatially resolved properties such as extinction, star formation rate and metallicity for this galaxy and our results are the following.

1. Our data shows an extinction map with maximum values located at the position of the main clumps of star formation showing small spatial variations. The colour excess values for the different clumps are in the range $\rm E(B-V)_{t}=0.12-0.21 mag$ (0.17 mag for the integrated galaxy spectrum), in agreement with the results from \cite{pasquali08}.

2. We derive a $\rm H\alpha$-based SFR for this galaxy of $6.2 \pm 0.8 \rm M_{\odot}/yr$ in agreement with the SFR derived from infrared and radio continuum fluxes \citep{garland05}. The star formation is located mainly in clumps A, B, C and F. Different properties measured in clump B makes it peculiar. We find for this burst the highest $\rm H\alpha$ luminosity with a SFR surface density of 0.5 $\rm M_{\odot}yr^{-1}kpc^{-2}$.
This region shows the absence of strong absorption features and the presence of Wolf-Rayet stellar population indicating this is a young burst of massive stars. 
The analysis of the emission line ratios discard the presence of any AGN activity or shocks as the ionization source in this clump or any other in this galaxy.

3. Furthermore, we estimate a gas metallicity of $\rm 12+log(O/H)=8.20\pm0.15$ (0.32 solar),  for the integrated galaxy using the R23 index. The values derived for the different clumps with this method show a small metallicity variations in this galaxy, with values in the range 8.12 (for clump A) - 8.23 (for clump B) for $\rm 12+log(O/H)$.

The detection ($5\sigma$-level) of the auroral forbidden emission line $\rm [OIII]\lambda4363$ in some fibres located in clump B allows us to determine metallicities using the Te-method and compare these values with the computed R23 oxygen abundance for the same fibres.
Although in some cases the R23 index overestimates the oxygen abundance (differences of 0.14-0.40 dex), in those fibres with higher signal-to-noise ratio in the $\rm [OIII]\lambda4363$ detection both methods give similar results suggesting the reliability of the R23 method.

4. Between the possible mechanisms to explain the starburst activity in this galaxy, our 3D spectroscopic data support the scenario in which a minor merger is taking place with the possibility for clump B, a compact decoupled kinematic component, to be a dwarf satellite galaxy.

\section*{Acknowledgments}
We are grateful to the anonymous referee for his valuable comments which have led to an improvement in this paper.
We acknowledge support from the Spanish Programa Nacional de Astronom{\'i}a y Astrof{\'i}sica under grants AYA2006-02358 and AYA2006-15698-C02-02. This work is partially funded by the Spanish MICINN under the Consolider-Ingenio 2010 Program grant CSD2006-00070: First Science with the GTC. ACM acknowledges support from the Juan de la Cierva Program financed by the Spanish MICINN.
JPG acknowledges support from a University of Florida Alumni Fellowship and RG from NASA Grant LTSA NA65-11635.
SFS thanks the sub-programs of {\it Viabilidad, Dise\~no, Acceso y Mejora de ICTS} ICTS-2009-10, and the {\it PAI Proyecto de Excelencia} P08-FWM-04319 and the funds of the PAI research group FQM360. SFS thanks the Spanish Plan Nacional de Astronom\'\i a program AYA2005-09413-C02-02.



\bibliographystyle{mn2e}
\bibliography{ref}

\begin{thebibliography}{}

\bibitem[\protect\citeauthoryear{{Adelman-McCarthy}, {Ag{\"u}eros} \&
  {Allam}}{{Adelman-McCarthy} et~al.}{2006}]{adelman06}
{Adelman-McCarthy} J.~K.,  {Ag{\"u}eros} M.~A.,    {Allam} e.~a.,  2006, ApJS,
  162, 38

\bibitem[\protect\citeauthoryear{{Allende Prieto}, {Lambert} \&
  {Asplund}}{{Allende Prieto} et~al.}{2001}]{allende01}
{Allende Prieto} C.,  {Lambert} D.~L.,    {Asplund} M.,  2001, ApJ, 556, L63

\bibitem[\protect\citeauthoryear{{Alloin}, {Collin-Souffrin}, {Joly} \&
  {Vigroux}}{{Alloin} et~al.}{1979}]{alloin79}
{Alloin} D.,  {Collin-Souffrin} S.,  {Joly} M.,    {Vigroux} L.,  1979, A\&A,
  78, 200

\bibitem[\protect\citeauthoryear{{Bell}}{{Bell}}{2003}]{bell03}
{Bell} E.~F.,  2003, ApJ, 586, 794

\bibitem[\protect\citeauthoryear{{Boselli}, {Gavazzi} \& {Sanvito}}{{Boselli}
  et~al.}{2003}]{boselli03}
{Boselli} A.,  {Gavazzi} G.,    {Sanvito} G.,  2003, A\&A, 402, 37

\bibitem[\protect\citeauthoryear{{Brinchmann}, {Charlot}, {White}, {Tremonti},
  {Kauffmann}, {Heckman} \& {Brinkmann}}{{Brinchmann}
  et~al.}{2004}]{brinchmann04}
{Brinchmann} J.,  {Charlot} S.,  {White} S.~D.~M.,  {Tremonti} C.,  {Kauffmann}
  G.,  {Heckman} T.,    {Brinkmann} J.,  2004, MNRAS, 351, 1151

\bibitem[\protect\citeauthoryear{{Calzetti}, {Armus}, {Bohlin}, {Kinney},
  {Koornneef} \& {Storchi-Bergmann}}{{Calzetti} et~al.}{2000}]{calzetti00}
{Calzetti} D.,  {Armus} L.,  {Bohlin} R.~C.,  {Kinney} A.~L.,  {Koornneef} J.,
    {Storchi-Bergmann} T.,  2000, ApJ, 533, 682

\bibitem[\protect\citeauthoryear{{Calzetti}, {Kinney} \&
  {Storchi-Bergmann}}{{Calzetti} et~al.}{1994}]{calzetti94}
{Calzetti} D.,  {Kinney} A.~L.,    {Storchi-Bergmann} T.,  1994, ApJ, 429, 582

\bibitem[\protect\citeauthoryear{{Cardelli}, {Clayton} \& {Mathis}}{{Cardelli}
  et~al.}{1989}]{cardelli89}
{Cardelli} J.~A.,  {Clayton} G.~C.,    {Mathis} J.~S.,  1989, ApJ, 345, 245

\bibitem[\protect\citeauthoryear{{Castellanos}}{{Castellanos}}{2000}]{castella%
nos00}
{Castellanos} M.,  2000, "Ph.D. Thesis, Universidad Autonoma de Madrid"

\bibitem[\protect\citeauthoryear{{Charlot} \& {Longhetti}}{{Charlot} \&
  {Longhetti}}{2001}]{charlot01}
{Charlot} S.,  {Longhetti} M.,  2001, MNRAS, 323, 887

\bibitem[\protect\citeauthoryear{{Condon} \& {Yin}}{{Condon} \&
  {Yin}}{1990}]{condon90}
{Condon} J.~J.,  {Yin} Q.~F.,  1990, ApJ, 357, 97

\bibitem[\protect\citeauthoryear{{Cortese}, {Boselli}, {Franzetti}, {Decarli},
  {Gavazzi}, {Boissier} \& {Buat}}{{Cortese} et~al.}{2008}]{cortese08}
{Cortese} L.,  {Boselli} A.,  {Franzetti} P.,  {Decarli} R.,  {Gavazzi} G.,
  {Boissier} S.,    {Buat} V.,  2008, MNRAS, 386, 1157

\bibitem[\protect\citeauthoryear{{Denicol{\'o}}, {Terlevich} \&
  {Terlevich}}{{Denicol{\'o}} et~al.}{2002}]{denicolo02}
{Denicol{\'o}} G.,  {Terlevich} R.,    {Terlevich} E.,  2002, MNRAS, 330, 69

\bibitem[\protect\citeauthoryear{{Dopita} \& {Evans}}{{Dopita} \&
  {Evans}}{1986}]{dopita86}
{Dopita} M.~A.,  {Evans} I.~N.,  1986, ApJ, 307, 431

\bibitem[\protect\citeauthoryear{{Duflot-Augarde} \& {Alloin}}{{Duflot-Augarde}
  \& {Alloin}}{1982}]{duflot82}
{Duflot-Augarde} R.,  {Alloin} D.,  1982, A\&A, 112, 257

\bibitem[\protect\citeauthoryear{{F{\"o}rster Schreiber}, {Genzel}, {Lehnert}
  \& {et al.}}{{F{\"o}rster Schreiber} et~al.}{2006}]{forster06}
{F{\"o}rster Schreiber} N.~M.,  {Genzel} R.,  {Lehnert} M.~D.,    {et al.}
  2006, ApJ, 645, 1062

\bibitem[\protect\citeauthoryear{{Gallego}, {Zamorano}, {Rego}, {Alonso} \&
  {Vitores}}{{Gallego} et~al.}{1996}]{gallego96}
{Gallego} J.,  {Zamorano} J.,  {Rego} M.,  {Alonso} O.,    {Vitores} A.~G.,
  1996, A\&AS, 120, 323

\bibitem[\protect\citeauthoryear{{Garland}, {Pisano}, {Williams}, {Guzm{\'a}n},
  {Castander} \& {Sage}}{{Garland} et~al.}{2007}]{garland07}
{Garland} C.~A.,  {Pisano} D.~J.,  {Williams} J.~P.,  {Guzm{\'a}n} R.,
  {Castander} F.~J.,    {Sage} L.~J.,  2007, ApJ, 671, 310

\bibitem[\protect\citeauthoryear{{Garland}, {Williams}, {Pisano}, {Guzm{\'a}n},
  {Castander} \& {Brinkmann}}{{Garland} et~al.}{2005}]{garland05}
{Garland} C.~A.,  {Williams} J.~P.,  {Pisano} D.~J.,  {Guzm{\'a}n} R.,
  {Castander} F.~J.,    {Brinkmann} J.,  2005, ApJ, 624, 714

\bibitem[\protect\citeauthoryear{{Gil de Paz}, {Boissier}, {Madore} \& {et
  al.}}{{Gil de Paz} et~al.}{2007}]{gildepaz07}
{Gil de Paz} A.,  {Boissier} S.,  {Madore} B.~F.,    {et al.} 2007, ApJS, 173,
  185

\bibitem[\protect\citeauthoryear{{Guzm{\'a}n}, {Gallego}, {Koo}, {Phillips},
  {Lowenthal}, {Faber}, {Illingworth} \& {Vogt}}{{Guzm{\'a}n}
  et~al.}{1997}]{guzman97}
{Guzm{\'a}n} R.,  {Gallego} J.,  {Koo} D.~C.,  {Phillips} A.~C.,  {Lowenthal}
  J.~D.,  {Faber} S.~M.,  {Illingworth} G.~D.,    {Vogt} N.~P.,  1997, ApJ,
  489, 559

\bibitem[\protect\citeauthoryear{{Guzm{\'a}n}, {Jangren}, {Koo}, {Bershady} \&
  {Simard}}{{Guzm{\'a}n} et~al.}{1998}]{guzman98}
{Guzm{\'a}n} R.,  {Jangren} A.,  {Koo} D.~C.,  {Bershady} M.~A.,    {Simard}
  L.,  1998, ApJ, 495, L13+

\bibitem[\protect\citeauthoryear{{Hammer}, {Gruel}, {Thuan}, {Flores} \&
  {Infante}}{{Hammer} et~al.}{2001}]{hammer01}
{Hammer} F.,  {Gruel} N.,  {Thuan} T.~X.,  {Flores} H.,    {Infante} L.,  2001,
  ApJ, 550, 570

\bibitem[\protect\citeauthoryear{{Heckman}, {Armus} \& {Miley}}{{Heckman}
  et~al.}{1990}]{heckman90}
{Heckman} T.~M.,  {Armus} L.,    {Miley} G.~K.,  1990, ApJS, 74, 833

\bibitem[\protect\citeauthoryear{{Homeier}, {Gallagher} III \&
  {Pasquali}}{{Homeier} et~al.}{2002}]{homeier02}
{Homeier} N.,  {Gallagher} III J.~S.,    {Pasquali} A.,  2002, A\&A, 391, 857

\bibitem[\protect\citeauthoryear{{Homeier} \& {Gallagher}}{{Homeier} \&
  {Gallagher}}{1999}]{homeier99}
{Homeier} N.~L.,  {Gallagher} J.~S.,  1999, ApJ, 522, 199

\bibitem[\protect\citeauthoryear{{Houck}, {Roellig}, {van Cleve}, {Forrest},
  {Herter}, {Lawrence}, {Matthews}, {Reitsema}, {Soifer}, {Watson}, {Weedman},
  {Huisjen} \& {Troeltzsch}}{{Houck} et~al.}{2004}]{houck04}
{Houck} J.~R.,  {Roellig} T.~L.,  {van Cleve} J.,  {Forrest} W.~J.,  {Herter}
  T.,  {Lawrence} C.~R.,  {Matthews} K.,  {Reitsema} H.~J.,  {Soifer} B.~T.,
  {Watson} D.~M.,  {Weedman} D.,  {Huisjen} M.,    {Troeltzsch} J.,  2004,
  ApJS, 154, 18

\bibitem[\protect\citeauthoryear{{Hoyos}, {Koo}, {Phillips}, {Willmer} \&
  {Guhathakurta}}{{Hoyos} et~al.}{2005}]{hoyos05}
{Hoyos} C.,  {Koo} D.~C.,  {Phillips} A.~C.,  {Willmer} C.~N.~A.,
  {Guhathakurta} P.,  2005, ApJ, 635, L21

\bibitem[\protect\citeauthoryear{{Huchra}}{{Huchra}}{1977}]{huchra77}
{Huchra} J.~P.,  1977, ApJS, 35, 171

\bibitem[\protect\citeauthoryear{{Huchra}, {Vogeley} \& {Geller}}{{Huchra}
  et~al.}{1999}]{huchra99}
{Huchra} J.~P.,  {Vogeley} M.~S.,    {Geller} M.~J.,  1999, ApJS, 121, 287

\bibitem[\protect\citeauthoryear{{Izotov}, {Thuan} \& {Lipovetsky}}{{Izotov}
  et~al.}{1994}]{izotov94}
{Izotov} Y.~I.,  {Thuan} T.~X.,    {Lipovetsky} V.~A.,  1994, ApJ, 435, 647

\bibitem[\protect\citeauthoryear{{Jog} \& {Das}}{{Jog} \& {Das}}{1992}]{jog92}
{Jog} C.~J.,  {Das} M.,  1992, ApJ, 400, 476

\bibitem[\protect\citeauthoryear{{Kelz}, {Verheijen}, {Roth}, {Bauer},
  {Becker}, {Paschke}, {Popow}, {S{\'a}nchez} \& {Laux}}{{Kelz}
  et~al.}{2006}]{kelz06}
{Kelz} A.,  {Verheijen} M.~A.~W.,  {Roth} M.~M.,  {Bauer} S.~M.,  {Becker} T.,
  {Paschke} J.,  {Popow} E.,  {S{\'a}nchez} S.~F.,    {Laux} U.,  2006, PASP,
  118, 129

\bibitem[\protect\citeauthoryear{{Kennicutt}
  Jr.}{{Kennicutt}}{1998a}]{kennicutt98}
{Kennicutt} Jr. R.~C.,  1998a, ARA\&A, 36, 189

\bibitem[\protect\citeauthoryear{{Kennicutt}
  Jr.}{{Kennicutt}}{1998b}]{kennicutt98a}
{Kennicutt} Jr. R.~C.,  1998b, ApJ, 498, 541

\bibitem[\protect\citeauthoryear{{Kewley} \& {Dopita}}{{Kewley} \&
  {Dopita}}{2002}]{kewley02}
{Kewley} L.~J.,  {Dopita} M.~A.,  2002, ApJS, 142, 35

\bibitem[\protect\citeauthoryear{{Kobulnicky} \& {Zaritsky}}{{Kobulnicky} \&
  {Zaritsky}}{1999}]{kobulnicky99}
{Kobulnicky} H.~A.,  {Zaritsky} D.,  1999, ApJ, 511, 118

\bibitem[\protect\citeauthoryear{{Koo}, {Bershady}, {Wirth}, {Stanford} \&
  {Majewski}}{{Koo} et~al.}{1994}]{koo94}
{Koo} D.~C.,  {Bershady} M.~A.,  {Wirth} G.~D.,  {Stanford} S.~A.,
  {Majewski} S.~R.,  1994, ApJ, 427, L9

\bibitem[\protect\citeauthoryear{{Kurucz}}{{Kurucz}}{1992}]{kurucz92}
{Kurucz} R.~L.,  1992, in Barbuy B., Renzini A., eds. Proc. IAU Symp. 149, 1,
  The stellar Populations of Galaxies. Kluwer Dordrecht, p. 225

\bibitem[\protect\citeauthoryear{{Lowenthal}, {Koo}, {Guzm{\'a}n}, {Gallego},
  {Phillips}, {Faber}, {Vogt}, {Illingworth} \& {Gronwall}}{{Lowenthal}
  et~al.}{1997}]{lowenthal97}
{Lowenthal} J.~D.,  {Koo} D.~C.,  {Guzm{\'a}n} R.,  {Gallego} J.,  {Phillips}
  A.~C.,  {Faber} S.~M.,  {Vogt} N.~P.,  {Illingworth} G.~D.,    {Gronwall} C.,
   1997, ApJ, 481, 673

\bibitem[\protect\citeauthoryear{{Maeder} \& {Conti}}{{Maeder} \&
  {Conti}}{1994}]{maeder94}
{Maeder} A.,  {Conti} P.~S.,  1994, ARA\&A, 32, 227

\bibitem[\protect\citeauthoryear{{Markarian}, {Lipovetsky}, {Stepanian},
  {Erastova} \& {Shapovalova}}{{Markarian} et~al.}{1989}]{markarian89}
{Markarian} B.~E.,  {Lipovetsky} V.~A.,  {Stepanian} J.~A.,  {Erastova} L.~K.,
    {Shapovalova} A.~I.,  1989, Soobshcheniya Spetsial'noj Astrofizicheskoj
  Observatorii, 62, 5

\bibitem[\protect\citeauthoryear{{McQuade}, {Calzetti} \& {Kinney}}{{McQuade}
  et~al.}{1995}]{mcquade95}
{McQuade} K.,  {Calzetti} D.,    {Kinney} A.~L.,  1995, ApJS, 97, 331

\bibitem[\protect\citeauthoryear{{Melbourne}, {Phillips}, {Harker}, {Novak},
  {Koo} \& {Faber}}{{Melbourne} et~al.}{2007}]{melbourne07}
{Melbourne} J.,  {Phillips} A.~C.,  {Harker} J.,  {Novak} G.,  {Koo} D.~C.,
  {Faber} S.~M.,  2007, ApJ, 660, 81

\bibitem[\protect\citeauthoryear{{Melnick}, {Moles}, {Terlevich} \&
  {Garcia-Pela\ yo}}{{Melnick} et~al.}{1987}]{melnick87}
{Melnick} J.,  {Moles} M.,  {Terlevich} R.,    {Garcia-Pela\ yo} J.,  1987,
  MNRAS, 226, 849

\bibitem[\protect\citeauthoryear{{Meurer}, {Heckman} \& {Calzetti}}{{Meurer}
  et~al.}{1999}]{meurer99}
{Meurer} G.~R.,  {Heckman} T.~M.,    {Calzetti} D.,  1999, ApJ, 521, 64

\bibitem[\protect\citeauthoryear{{Mulder} \& {van Driel}}{{Mulder} \& {van
  Driel}}{1996}]{mulder96}
{Mulder} P.~S.,  {van Driel} W.,  1996, A\&A, 309, 403

\bibitem[\protect\citeauthoryear{{Noeske}, {Koo}, {Phillips}, {Willmer},
  {Melbourne}, {Gil de Paz} \& {Papaderos}}{{Noeske} et~al.}{2006}]{noeske06}
{Noeske} K.~G.,  {Koo} D.~C.,  {Phillips} A.~C.,  {Willmer} C.~N.~A.,
  {Melbourne} J.,  {Gil de Paz} A.,    {Papaderos} P.,  2006, ApJ, 640, L143

\bibitem[\protect\citeauthoryear{{Osterbrock}}{{Osterbrock}}{1989}]{osterbrock%
89}
{Osterbrock} D.~E.,  1989, {Astrophysics of gaseous nebulae and active galactic
  nuclei}.
Research supported by the University of California, John Simon Guggenheim
  Memorial Foundation, University of Minnesota, et al.~Mill Valley, CA,
  University Science Books, 1989, 422 p.

\bibitem[\protect\citeauthoryear{{Pagel}, {Edmunds}, {Blackwell}, {Chun} \&
  {Smith}}{{Pagel} et~al.}{1979}]{pagel79}
{Pagel} B.~E.~J.,  {Edmunds} M.~G.,  {Blackwell} D.~E.,  {Chun} M.~S.,
  {Smith} G.,  1979, MNRAS, 189, 95

\bibitem[\protect\citeauthoryear{{Pagel}, {Simonson}, {Terlevich} \&
  {Edmunds}}{{Pagel} et~al.}{1992}]{pagel92}
{Pagel} B.~E.~J.,  {Simonson} E.~A.,  {Terlevich} R.~J.,    {Edmunds} M.~G.,
  1992, MNRAS, 255, 325

\bibitem[\protect\citeauthoryear{{Pasquali} \& {Castangia}}{{Pasquali} \&
  {Castangia}}{2008}]{pasquali08}
{Pasquali} A.,  {Castangia} P.,  2008, MNRAS, 385, 468

\bibitem[\protect\citeauthoryear{{P{\'e}rez-Gallego}, {Guzm{\'a}n},
  {Castillo-Morales}, {Castander}, {Gallego}, {Garland}, {Gruel}, {Pisano},
  {S{\'a}nchez} \& {Zamorano}}{{P{\'e}rez-Gallego}
  et~al.}{2010}]{perez-gallego10}
{P{\'e}rez-Gallego} J.,  {Guzm{\'a}n} R.,  {Castillo-Morales} A.,  {Castander}
  F.~J.,  {Gallego} J.,  {Garland} C.~A.,  {Gruel} N.,  {Pisano} D.~J.,
  {S{\'a}nchez} S.~F.,    {Zamorano} J.,  2010, MNRAS, 402, 1397

\bibitem[\protect\citeauthoryear{{P{\'e}rez-Gonz{\'a}lez}, {Zamorano},
  {Gallego}, {Arag{\'o}n-Salamanca} \& {Gil de Paz}}{{P{\'e}rez-Gonz{\'a}lez}
  et~al.}{2003}]{perez-gonzalez03}
{P{\'e}rez-Gonz{\'a}lez} P.~G.,  {Zamorano} J.,  {Gallego} J.,
  {Arag{\'o}n-Salamanca} A.,    {Gil de Paz} A.,  2003, ApJ, 591, 827

\bibitem[\protect\citeauthoryear{{P{\'e}rez-Gonz{\'a}lez}, {Zamorano},
  {Gallego} \& {Gil de Paz}}{{P{\'e}rez-Gonz{\'a}lez}
  et~al.}{2000}]{perez-gonzalez00}
{P{\'e}rez-Gonz{\'a}lez} P.~G.,  {Zamorano} J.,  {Gallego} J.,    {Gil de Paz}
  A.,  2000, A\&AS, 141, 409

\bibitem[\protect\citeauthoryear{{Phillips}, {Guzm{\'a}n}, {Gallego}, {Koo},
  {Lowenthal}, {Vogt}, {Faber} \& {Illingworth}}{{Phillips}
  et~al.}{1997}]{phillips97}
{Phillips} A.~C.,  {Guzm{\'a}n} R.,  {Gallego} J.,  {Koo} D.~C.,  {Lowenthal}
  J.~D.,  {Vogt} N.~P.,  {Faber} S.~M.,    {Illingworth} G.~D.,  1997, ApJ,
  489, 543

\bibitem[\protect\citeauthoryear{{Pisano}, {Kobulnicky}, {Guzm{\'a}n},
  {Gallego} \& {Bershady}}{{Pisano} et~al.}{2001}]{pisano01}
{Pisano} D.~J.,  {Kobulnicky} H.~A.,  {Guzm{\'a}n} R.,  {Gallego} J.,
  {Bershady} M.~A.,  2001, AJ, 122, 1194

\bibitem[\protect\citeauthoryear{{Puech}, {Hammer}, {Flores}, {{\"O}stlin} \&
  {Marquart}}{{Puech} et~al.}{2006}]{puech06}
{Puech} M.,  {Hammer} F.,  {Flores} H.,  {{\"O}stlin} G.,    {Marquart} T.,
  2006, A\&A, 455, 119

\bibitem[\protect\citeauthoryear{{Richer} \& {McCall}}{{Richer} \&
  {McCall}}{1995}]{richer95}
{Richer} M.~G.,  {McCall} M.~L.,  1995, ApJ, 445, 642

\bibitem[\protect\citeauthoryear{{Rickes}, {Pastoriza} \& {Bonatto}}{{Rickes}
  et~al.}{2008}]{rickes08}
{Rickes} M.~G.,  {Pastoriza} M.~G.,    {Bonatto} C.,  2008, MNRAS, 384, 1427

\bibitem[\protect\citeauthoryear{{Salzer}, {Williams} \& {Gronwall}}{{Salzer}
  et~al.}{2009}]{salzer09}
{Salzer} J.~J.,  {Williams} A.~L.,    {Gronwall} C.,  2009, ApJ, 695, L67

\bibitem[\protect\citeauthoryear{{S{\'a}nchez}}{{S{\'a}nchez}}{2004}]{sanchez0%
4}
{S{\'a}nchez} S.~F.,  2004, Astronomische Nachrichten, 325, 167

\bibitem[\protect\citeauthoryear{{S{\'a}nchez}}{{S{\'a}nchez}}{2006}]{sanchez0%
6}
{S{\'a}nchez} S.~F.,  2006, Astronomische Nachrichten, 327, 850

\bibitem[\protect\citeauthoryear{{Sanders} \& {Mirabel}}{{Sanders} \&
  {Mirabel}}{1996}]{sanders96}
{Sanders} D.~B.,  {Mirabel} I.~F.,  1996, ARA\&A, 34, 749

\bibitem[\protect\citeauthoryear{{Schlegel}, {Finkbeiner} \&
  {Davis}}{{Schlegel} et~al.}{1998}]{schlegel98}
{Schlegel} D.~J.,  {Finkbeiner} D.~P.,    {Davis} M.,  1998, ApJ, 500, 525

\bibitem[\protect\citeauthoryear{{Schmitt}, {Calzetti}, {Armus}, {Giavalisco},
  {Heckman}, {Kennicutt} Jr., {Leitherer} \& {Meurer}}{{Schmitt}
  et~al.}{2006}]{schmitt06}
{Schmitt} H.~R.,  {Calzetti} D.,  {Armus} L.,  {Giavalisco} M.,  {Heckman}
  T.~M.,  {Kennicutt} Jr. R.~C.,  {Leitherer} C.,    {Meurer} G.~R.,  2006,
  ApJ, 643, 173

\bibitem[\protect\citeauthoryear{{Schweizer}}{{Schweizer}}{1987}]{schweizer87}
{Schweizer} F.,  1987, in {S.~M.~Faber} ed., Nearly Normal Galaxies. From the
  Planck Time to the Present {Star formation in colliding and merging
  galaxies}.
pp 18--25

\bibitem[\protect\citeauthoryear{{Shlosman}, {Begelman} \& {Frank}}{{Shlosman}
  et~al.}{1990}]{shlosman90}
{Shlosman} I.,  {Begelman} M.~C.,    {Frank} J.,  1990, Nat, 345, 679

\bibitem[\protect\citeauthoryear{{Smith}, {Armus}, {Dale}, {Roussel}, {Sheth},
  {Buckalew}, {Jarrett}, {Helou} \& {Kennicutt} Jr.}{{Smith}
  et~al.}{2007}]{smith07}
{Smith} J.~D.~T.,  {Armus} L.,  {Dale} D.~A.,  {Roussel} H.,  {Sheth} K.,
  {Buckalew} B.~A.,  {Jarrett} T.~H.,  {Helou} G.,    {Kennicutt} Jr. R.~C.,
  2007, PASP, 119, 1133

\bibitem[\protect\citeauthoryear{{Stasi{\'n}ska}}{{Stasi{\'n}ska}}{1990}]{stas%
inska90}
{Stasi{\'n}ska} G.,  1990, A\&AS, 83, 501

\bibitem[\protect\citeauthoryear{{Steidel}, {Adelberger}, {Shapley}, {Pettini},
  {Dickinson} \& {Giavalisco}}{{Steidel} et~al.}{2003}]{steidel03}
{Steidel} C.~C.,  {Adelberger} K.~L.,  {Shapley} A.~E.,  {Pettini} M.,
  {Dickinson} M.,    {Giavalisco} M.,  2003, ApJ, 592, 728

\bibitem[\protect\citeauthoryear{{Steidel}, {Giavalisco}, {Pettini},
  {Dickinson} \& {Adelberger}}{{Steidel} et~al.}{1996}]{steidel96}
{Steidel} C.~C.,  {Giavalisco} M.,  {Pettini} M.,  {Dickinson} M.,
  {Adelberger} K.~L.,  1996, ApJ, 462, L17+

\bibitem[\protect\citeauthoryear{{Storchi-Bergmann}, {Calzetti} \&
  {Kinney}}{{Storchi-Bergmann} et~al.}{1994}]{storchi94}
{Storchi-Bergmann} T.,  {Calzetti} D.,    {Kinney} A.~L.,  1994, ApJ, 429, 572

\bibitem[\protect\citeauthoryear{{Weedman}}{{Weedman}}{1983}]{weedman83}
{Weedman} D.~W.,  1983, ApJ, 266, 479

\bibitem[\protect\citeauthoryear{{Weedman} \& {Houck}}{{Weedman} \&
  {Houck}}{2009}]{weedman09}
{Weedman} D.~W.,  {Houck} J.~R.,  2009, ApJ, 693, 370

\bibitem[\protect\citeauthoryear{{Werk}, {Jangren} \& {Salzer}}{{Werk}
  et~al.}{2004}]{werk04}
{Werk} J.~K.,  {Jangren} A.,    {Salzer} J.~J.,  2004, ApJ, 617, 1004

\bibitem[\protect\citeauthoryear{{Werner}, {Roellig}, {Low}, {Rieke}, {Rieke},
  {Hoffmann}, {Young} \& {Houck}}{{Werner} et~al.}{2004}]{werner04}
{Werner} M.~W.,  {Roellig} T.~L.,  {Low} F.~J.,  {Rieke} G.~H.,  {Rieke} M.,
  {Hoffmann} W.~F.,  {Young} E.,    {Houck} J.~R.,  2004, ApJS, 154, 1

\bibitem[\protect\citeauthoryear{{Zamorano}, {Rego}, {Gallego}, {Vitores},
  {Gonzalez-Riestra} \& {Rodriguez-Caderot}}{{Zamorano}
  et~al.}{1994}]{zamorano94}
{Zamorano} J.,  {Rego} M.,  {Gallego} J.,  {Vitores} A.~G.,  {Gonzalez-Riestra}
  R.,    {Rodriguez-Caderot} G.,  1994, ApJS, 95, 387

\end{thebibliography}

\end{document}